\documentstyle[subequation,epsfig]{article}
\def\circa#1{\,\raise.3ex\hbox{$#1$\kern-.75em\lower1ex\hbox{$\sim$}}\,}

\newcommand  \f  \varphi
\newcommand{\dl}{L_W(s)}
\newcommand \bra {\langle}
\newcommand \ket {\rangle}
\newcommand{\be}{\begin{equation}}
\newcommand{\ee}{\end{equation}}
\newcommand{\ben}{\begin{displaymath}}
\newcommand{\een}{\end{displaymath}}
\newcommand{\ba}{\begin{eqnarray}}
\newcommand{\ea}{\end{eqnarray}}
\newcommand{\ban}{\begin{eqnarray*}}
\newcommand{\ean}{\end{eqnarray*}}
\newcommand{\cro}{\dagger}

\newcommand{\de}{\partial}
\newcommand{\tvet}{\mbox{\boldmath $t$}}

\newcommand{\kvet}{\mbox{\boldmath $k$}}
\newcommand{\Kvet}{\mbox{\boldmath $K$}}
\newcommand{\pvet}{\mbox{\boldmath $p$}}

\begin{document}
\begin{titlepage}
\vspace{1cm}

\begin{center}
{\huge \bf Electroweak}
\\

{\huge \bf Bloch-Nordsieck violation at the TeV scale:}
\\
\vskip.2cm
{\huge \bf ``strong'' weak interactions ?}

\vskip1.5cm

{\large Marcello Ciafaloni}

{\it Dipartimento di Fisica, Universit\`a di Firenze e
\\ INFN - Sezione di Firenze,
 I-50125 Florence , Italy\\
E-mail: ciafaloni@fi.infn.it}

\vskip.5cm

{\large Paolo Ciafaloni}

{\it INFN - Sezione di Lecce,
\\Via per Arnesano, I-73100 Lecce, Italy
\\ E-mail: Paolo.Ciafaloni@le.infn.it}

\vskip.5cm

{\large Denis Comelli}

{\it INFN - Sezione di Ferrara,
\\Via Paradiso 12, I-35131 Ferrara, Italy\\
E-mail: comelli@fe.infn.it}

\vspace{4.5cm}
{\large\bf Abstract}
\end{center}
\begin{quotation}
Hard processes at the TeV scale exhibit enhanced (double log) EW  corrections
even for inclusive observables, leading to violation of the Bloch-Nordsieck
 theorem.
This effect, previously related to the non abelian nature of free 
 EW charges in the initial state ($e^- e^+$, $e^- p$, $p {p}$ ...),
is here investigated for fermion initiated hard processes and to all orders 
in EW
couplings. We find that the effect is important, especially for lepton
initiated processes, producing weak effects
that in some cases compete in magnitude with the strong ones.
We show that this  (double log) BN violating effect has
a universal energy dependence, related to the Sudakov
form factor in the adjoint representation.
The role of this form factor is to suppress cross section differences within
a weak isospin doublet, so that  at very large energy 
the cross sections for left-handed {\it electron}-positron and  
{\it neutrino}-positron scattering become equal.
Finally, we briefly discuss the phenomenological relevance of our results 
for future colliders.
\end{quotation}

\vspace{3cm}
\end{titlepage}

\def\baselinestretch{1.1}

\vskip1.3cm

\section{Introduction}

Enhanced electroweak corrections at the TeV scale have been recently
investigated by various authors \cite{cc1}-\cite{fadin} starting from the 
observation, made by
two of us \cite{cc1}, that double and single logarithms of
Sudakov type are present and sizeable in fixed angle fermion antifermion
annihilation processes at NLC energies. These effects produce, namely,
energy-growing corrections $\propto \alpha_W \log^2\frac{s}{M^2}$, the weak
scale $M\sim$ 90 GeV providing a physical cutoff for infrared and collinear
divergences. 

In a recent paper \cite{3p} we have pointed out a different but related
effect, which is peculiar of electroweak interactions.  
Due to the non abelian nature of electroweak charges
in the initial state, the double logs 
persist at {\it inclusive} level, thus leading to violation of the
Bloch-Nordsieck theorem \cite{bl}, in the sense that the dependence
on the IR cutoff $M$ is not washed out when summing real
and virtual corrections, as is usually the case.
The peculiar aspect which makes such double logs observable is symmetry
breaking itself, which generates the physical cutoff $M$ on one hand, and
allows the preparation of initial states as free abelian charges on the 
other\footnote{For instance, an electron is prepared at low energy $\ll M$ as
an abelian (QED) charge. As it is accelerated at energies $\gg M$, its charge
acquires a fully nonabelian character.}.
In fact, the BN theorem is in principle violated in QCD also 
\cite{dft}-\cite{col}, but in such case confinement forces a color averaging 
in the
initial state, which washes out the effect eventually. 
The ``preconfinement''
 features, pointed out at various stages \cite{ccm,av} 
mean that free quark asymptotic states make no sense, even at perturbative
level, because of form factors analogous to the ones discussed here.
The situation is different of course in the EW case: the analogous of color
averaging would mean for instance averaging over the cross
sections for $\nu e^+$ and $e^-e^+$; this 
 is meaningless from an experimental point
of view.

In this paper we investigate the structure of double log EW corrections to
all orders for light fermion initiated hard processes,  and we  characterize 
them by a universal energy dependence,
related to the EW Sudakov form factor in the adjoint representation.
The core of such analysis is provided by two lines of thought.
One, explained in Sec.3, is  based on the observation that only W
contributions are BN violating, the $Z$ and $\gamma$ ones being canceled
between virtual and real emission terms.
The energy dependence is then universal because the W couples universally to
left handed fermions and can be calculated to all orders by a simple Feynman
diagram technique.

The other line of thought, explained in Sec.5 following the coherent state
formalism of Sec.4, is based on the isospin
structure of the overlap matrix, describing the squared matrix element.
With a proper choice of indices, the singlet projection provides the isospin 
average (i.e. for instance $\sigma_{\nu e^+}+\sigma_{e^- e^+}$), 
and the vector projection the cross sections difference (i.e. 
$\sigma_{\nu e^+}-\sigma_{e^- e^+}$).
Only the latter is suppressed by a Sudakov form factor, which refers to the
adjoint representation, and is thus universal.
The latter approach is made completely rigorous by the use of the general
coherent state formalism \cite{fk,cm}
introduced for cancellation theorems in Sec.2, and by the important
observation that the photon scale plays no role for the (inclusive)
 BN violating terms.
This fact follows from the explicit cancellation of $Z$ and $\gamma$
contributions found by the diagrammatic approach, so that the effect starts,
and the symmetry is effectively restored at, the same $W$ threshold $M$.

Applications to physical processes are analyzed in Sec.6 and discussed in 
Sec.7.
We limit ourselves to lepton and/or quark large angle scattering as trigger
process and we classify the results according to the initial state, which is
provided by the accelerator.
Important corrections are found, especially for lepton initiated processes, 
as in the case of
$e\bar{e}$ (NLC) or $e p$ and $\bar{e}p$.              
Roughly speaking, we find in this case a 50 percent reduction in hadron beams 
compared to lepton beams, mostly due to the hadrons acting as weak isospin 
mixtures in the initial state.
Another important feature is the strong dependence of the
BN violating effects on the polarization of the colliding beams, of particular
relevance for NLCs \cite{NLC}.

\section{Cancellation theorems and Bloch-Nordsieck violation}

We consider the
structure of soft interactions accompanying a hard SM process, of type
\be\label{eq:1}
\{\alpha^I_1p^I_1,\alpha^I_2p^I_2\}\rightarrow
\{\alpha^F_1p^F_1,\alpha^F_2p^F_2,\dots,\alpha^F_np^F_n\}
\ee
where $\alpha,p$ denote isospin/color and  momentum indices of the initial and
final states, that we collectively denote by $\{\alpha_Ip_I\}$
and $\{\alpha_Fp_F\}$. The S matrix for such a process can be written
as an operator in the soft Hilbert space ${\cal H}_S$, that collects
the states which are almost degenerate with the hard ones, in the
form
\be\label{eq:2}
S={\cal U}^F_{\alpha_F\beta_F}\! (a_s,a^\cro_s)\;\;\;
S^H_{\beta_F\beta_I}\! (p_F,p_I)\;\;\;
{\cal U}^I_{\alpha_I\beta_I}\! (a_s,a^\cro_s)
\ee
where ${\cal U}^F$ and ${\cal U}^I$ are operator functionals of the
soft emission operators $a_s,a^\cro_s$.

Eq. (\ref{eq:2}) is supposed to be of general validity \cite{lg,cm},
because it rests
essentially on the separation of long-time interactions (the initial
and final ones described by the ${\cal U}$'s), and the short-time
hard interaction, described by $S^H$. The real problem is to find the
form of the ${\cal U}$'s , which is well known in QED \cite{fk}, has
been widely investigated 
in QCD \cite{cm}, and is under debate in the electroweak case
\cite{cc2} -\cite{fadin}.
 Their only general property is unitarity in the soft Hilbert
space ${\cal H}_S$, i.e.
\be\label{eq:3}
{\cal U}_{\alpha\beta}{\cal U}^\cro_{\beta\alpha'}=
{\cal U}^\cro_{\alpha\beta}{\cal U}_{\beta\alpha'}=
\delta_{\alpha\alpha'}
\ee

The key cancellation theorem satisfied by Eq. (\ref{eq:2}) is due to
Lee, Nauenberg and Kinoshita \cite{kln}, and states that soft
singularities cancel upon summation over initial and final soft states
which are degenerate with the hard ones:
\be
\sum_{i\in \Delta(p_I)}^{f\in \Delta(p_F)}
|\bra f|S|i\ket|^2={\rm Tr}_{{\cal H}_S}
({\cal U}^{I\cro}S^{H\cro}{\cal U}^{F\cro}
{\cal U}^{F} S^H {\cal U}^I)={\rm Tr}_{\alpha_I}
(S^{H\cro}(p_F,p_I)S^H(p_F,p_I))
\ee
where $\Delta(p_I,p_F)$ denote the sets of such soft states, and we
have used the unitarity property (\ref{eq:3}).

Although general, the KLN theorem is hardly of direct use, because it
involves the sum over the initial degenerate set, which is not
available experimentally. In the QED case, however, there is only an
abelian charge index, so that ${\cal U}_I$ commutes with
$S^{H\cro}S^H$, and cancels out by sum over the final degenerate set only.
This is the BN theorem: observables which are inclusive
over soft final states are infrared safe.

If the theory is non abelian, like QCD or the electroweak one under 
consideration, the BN theorem is generally violated,
because the initial state interaction is not canceled, i.e., by working in
color space, 
\be\label{eq:5}
\sum_{f\in \Delta(p_F)}|\bra f|S|i\ket|^2=
\;_S\bra 0|
{\cal U}^{I\cro}_{\alpha_I\beta'_I}(S^{H\cro}S^H)_{\beta'_I\beta_I}
{\cal U}^I_{\beta_I\alpha_I}|0\ket_S
=(S^{H\cro}S^H)_{\alpha_I\alpha_I}
+\Delta\sigma_{\alpha_I}
\ee 
where the ${\alpha_I}$ indices are not summed over, and 
$\Delta\sigma_{\alpha_I}$ is, in general, nonvanishing and IR singular.\\
Fortunately, in QCD the BN cancellation is essentially recovered
because of two features: (i) the need of initial color averaging,
because hadrons are colorless, and (ii) the commutativity of the
leading order coherent state operators (${\cal U}^l$)
 for any given color indices \cite{cm}:
\be\label{eq:6}
{\cal U}^l={\cal U}^l\! (a_s-a_s^\cro )\quad ,\quad
[{\cal U}^l_{\alpha\beta},{\cal U}^l_{\alpha'\beta'}]=0
\ee
We obtain therefore
\be
\sum_{color}{\cal U}^{l \cro}_{\alpha_I\beta'_I}
(S^{H\cro}S^H)_{\beta'_I\beta_I}
{\cal U}^l_{\beta_I\alpha_I}=
\sum_{color}(S^{H\cro}S^H)_{\beta'_I\beta_I} {\cal U}^l_{\beta_I\alpha_I}
{\cal U}^{l \cro}_{\alpha_I\beta'_I}
=
 Tr_{color}
S^{H\cro}S^H
\ee
thus recovering an infrared safe result (for subleading
features, see Refs. \cite{lrs,col,ccm}). 

In the electroweak case, in which
$M$ provides the physical infrared cutoff, there is no way out,
because the initial state is prepared with a fixed non abelian
charge. Therefore Eq. (\ref{eq:5}) applies, and double log
corrections $\sim \alpha_W\log^2\frac{s}{M^2}$
must affect any
observable associated with a hard process, even  the ones which are
inclusive over final soft bosons. This fact is surprising, because one
would expect such observables to depend only on energy and on
running couplings, while the double logs represent an
explicit $M$ (infrared cutoff) dependence. 
\section{Lowest order calculation and picture of higher orders}

In order to compute the uncanceled double logs, we first notice that,
according to Eq. (\ref{eq:5}), only initial state interactions need to be taken
into account. Here we give a diagrammatic account of the calculations, by
considering EW corrections to the overlap
matrix $O_H\equiv S_H^\cro S_H$, in which soft bosonic lines are emitted and/or
absorbed on initial lines only. In the following sections we shall give a more
rigorous treatment, based on the coherent state approach  that we have
introduced previously.

\begin{figure}[htb]\setlength{\unitlength}{1cm}
\begin{picture}(12,7.5)
\put(0.3,1){\epsfig{file=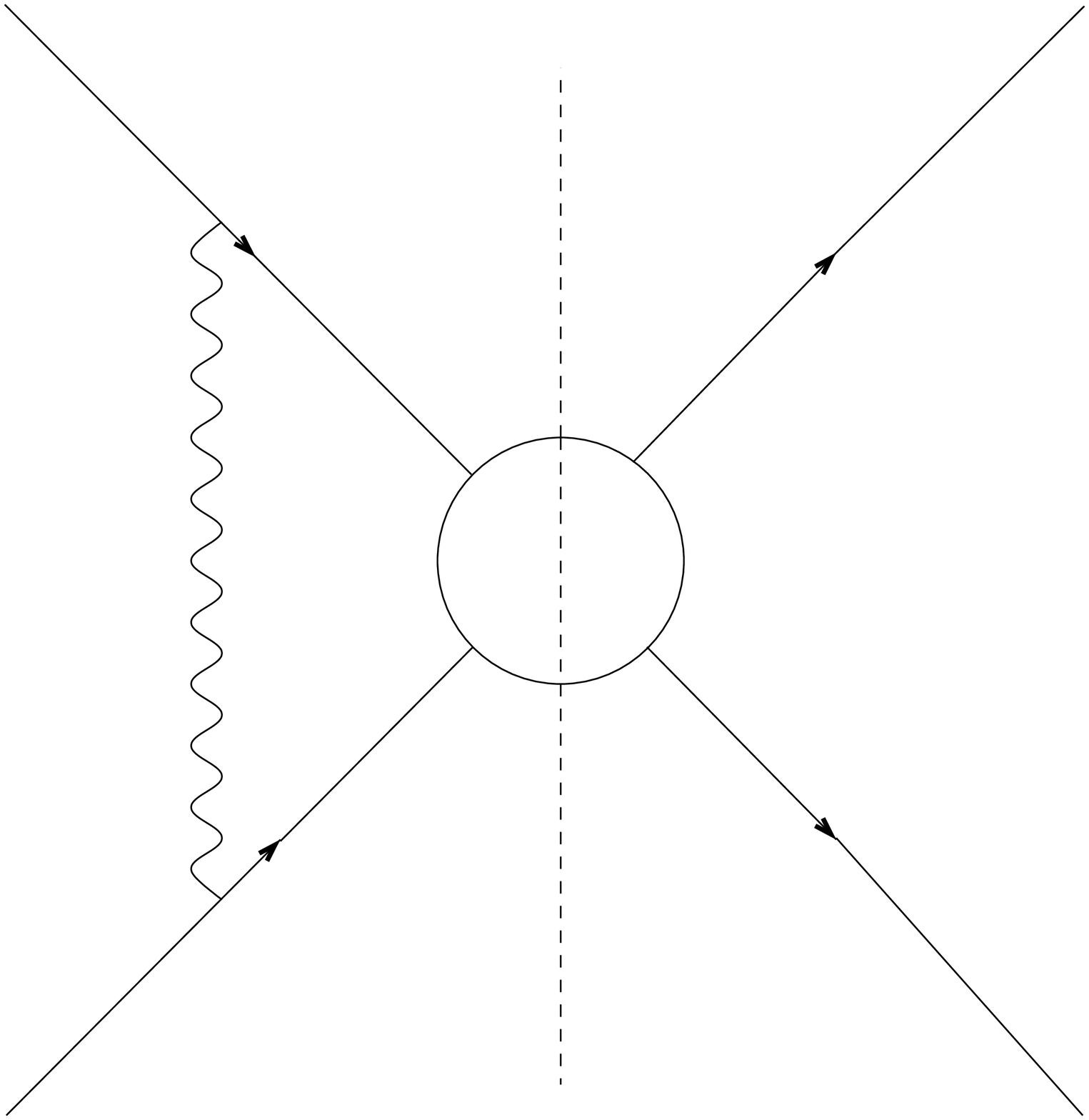,height=6cm}}
\put(.6,4){$k,a$}
\put(.7,6.8){$p_2,\alpha_2$}
\put(.7,0.9){$p_1,\alpha_1$}
\put(3.2,0.52){(a)}\put(12.2,.52){(b)}
\put(10.1,2.2){$t_1^a$}\put(13.7,6.1){$t_2'^a$}
\put(14.3,4){$k,a$}
\put(9,6.8){2}\put(15.3,6.8){2}
\put(9,0.9){1}\put(15.3,0.9){1}
\put(6.4,6.8){$\beta_2$}\put(6.4,0.9){$\beta_1$}
\put(2.8,3.9){$S_H^\cro\,\, S_H$}
\put(9.3,1){\epsfig{file=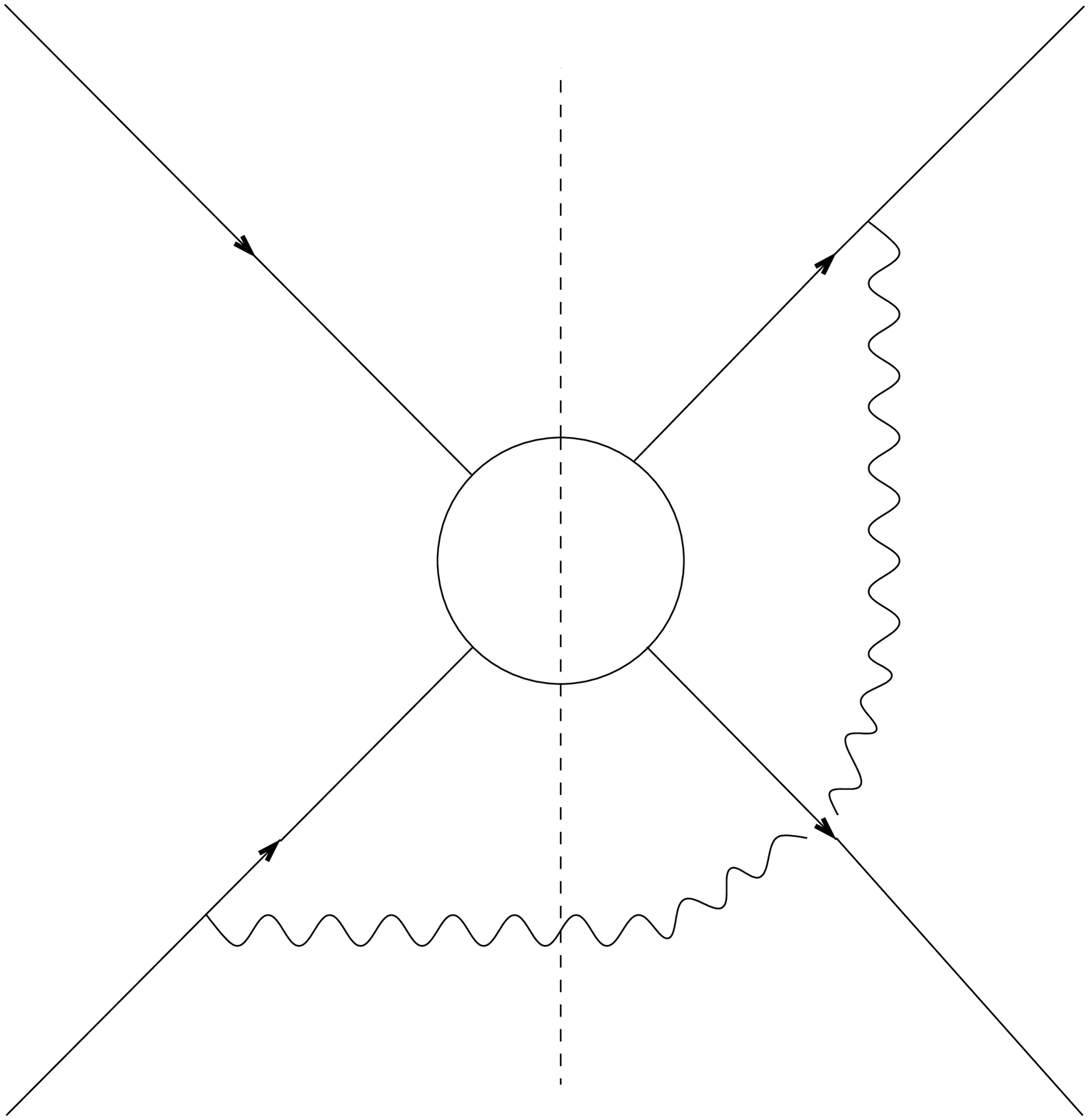,height=6cm}}
\put(11.8,3.9){$S_H^\cro\,\, S_H$}
\end{picture}
\caption{\label{1loopfig}
Unitarity diagrams for (a) virtual and (b) real emission
 contributions to lowest order initial state interactions 
in the Feynman gauge. Sum over gauge bosons a= $\gamma,Z,W$ and over
permutations is understood.}
\end{figure}

We start treating the lowest order soft EW contributions
to $\Delta\sigma\equiv\sigma-\sigma^H$ for a basic hard process
involving a hard scale $E\gg M_W\sim M_Z\equiv M$, and two  
massless fermions. We consider here the case for two L initial fermions,
both carrying  
nonabelian isospin (SU(2)) indices; a more general treatment
is demanded to section 5.
Since we only consider inclusive processes,
a sum over degenerate  final
states is understood, and we drop the superscript indicating initial states:
$\alpha_i^I\to\alpha_i$. In isospin space, the hard cross section structure
is then defined by the so called
hard overlap matrix, describing the squared matrix element: 
$\bra \beta_1 \beta_2| S_H^\cro S_H|\alpha_1\alpha_2\ket\equiv 
(O_H)_{\beta_1\beta_2,\alpha_1\alpha_2}$ (see Fig. \ref{1loopfig}).
While for cross sections we always have
$\alpha_i=\beta_i$, we leave open the possibility 
that $\alpha_i\neq\beta_i$ and see $O_H$ as an operator in isospin
space with four indices.

Since pure
QED corrections, at energies below $M$, cancel out automatically as noticed
before, we limit ourselves to bosonic energies $M\ll w\ll E$, for which the
gauge bosons $\gamma,Z,W$ have all approximately the same momenta, with $M$
acting as the only IR cutoff. This amounts eventually to setting the effective
photon scale $\lambda=M$, and neglecting symmetry breaking effects at the
inclusive double log level we are working. We would like to stress again
the fact
that this holds only for completely inclusive quantities, while for
(partially) exclusive observables the presence of a new scale $\lambda\neq M$
is unavoidable and may lead to symmetry breaking effects, a topic which is
currently under discussion \cite{cc2,fadin}.

At lowest order, the above assumption is easily verified, because
in the  hard (Born)
matrix element, since we work in a limit in which all kinematical
invariants are much bigger than gauge bosons
masses: $|s|,|t|,|u|\gg M^2$, the full gauge symmetry
SU(3)$\otimes$SU(2)$\otimes$U(1) is restored. Boson emission and
absorption is then described by the external (initial) line insertions of the
eikonal currents 
\be\label{correnti}
J_a^\mu=g[\frac{p_1^\mu}{kp_1}(t_1^a-t_1'^a)+
\frac{p_2^\mu}{kp_2}(t_2^a-t_2'^a)]
\qquad
J_0^\mu=g'[\frac{p_1^\mu}{kp_1}(Y_1-Y_1')+\frac{p_2^\mu}{kp_2}(Y_2-Y_2')]
\ee
Here a=1,2,3 is the SU(2) index, we work in the unbroken basis
$A_0=c_W \gamma-s_W Z,A_3=s_WA+c_WZ$, $g$ and $g'$ being the usual 
electroweak couplings with $\frac{s_W}{c_W}=\frac{g'}{g}$. 
The isospin operator $t_1(t_1')$ acts on the $\alpha_1$ $(\beta_1)$
index, and so on.
The currents (\ref{correnti}) are conserved $(k\cdot J=0)$ because of charge 
conservation at Born level:
\be\label{isocon}
t_1^a+t_2^a=t_1'^a+t_2'^a\quad ,\quad Y_1+Y_2=Y_1'+Y_2'
\ee
Since both $Y$ and $t_3$ are diagonal matrices, the $A_0,A_3$ (or $\gamma,Z$) 
contributions to the cross section {\it cancel out
automatically} between virtual and real emission terms ($t_i^3=t_i'^3,
Y_i=Y_i'$). The $W$
contributions are instead provided by:
\be\label{10}
g^2\frac{2p_1p_2}{(kp_1)(kp_2)}(\tvet_1-\tvet_1')\cdot (\tvet_2-\tvet_2')
\qquad\quad (\tvet_1\cdot\tvet_2\equiv\sum_a t_1^at_2^a)
\ee
where we have kept, for notational convenience, the vanishing $A_3$
contribution, and the charge factor can be replaced, because of the 
conservation (\ref{isocon}), by
\be
(\tvet_1-\tvet_1')\cdot (\tvet_2-\tvet_2')=
-(\tvet_1-\tvet_1')^2=2\tvet_1\cdot\tvet_1'-\tvet_1^2-\tvet_1'^2
\ee
The latter expression provides the charge computation in the axial
gauge, because in this gauge  the $W$ emission and absorption takes place 
on the same leg,
for both virtual ($-2\tvet_1^2$) and real emission ($2\tvet_1\cdot\tvet_1'$)
contributions. We can see from Fig. \ref{1loopfig}
the reason for noncancellation between
virtual 
and real one loop corrections:
the crucial point is that while  $\gamma, Z$ emission does not change the
initial state, $W$ emission does. Then, in the $W$ case virtual corrections
(Fig. \ref{1loopfig}a) are  proportional to $\sigma_{e\bar{e}}$, while
real corrections (Fig. \ref{1loopfig}b)
are of opposite sign but 
proportional to $ \sigma_{\nu\bar{e}}\neq \sigma_{e\bar{e}}$, 
giving rise to a 
non complete cancellation.

In order to simplify our considerations, let us note that, because of isospin
conservation, we must have (see (\ref{43}))
\be
\sigma_{\nu\bar{\nu}}=\sigma_{e\bar{e}}\quad ,\quad
\sigma_{\bar{\nu}e}=\sigma_{\nu\bar{e}}
\ee
We can therefore limit ourselves to the cross sections
$\sigma_\alpha\equiv\sigma_{\alpha\bar{e}}$ where $\alpha=\nu,e$ is a single
isospin index. Since real $W$ emission changes the isospin index, while virtual
corrections don't, we can summarize the first order calculations based on
(\ref{correnti}) and (\ref{10}) in the form
\be\label{firstorder}
\Delta\sigma_\alpha^1=\dl [-\delta_{\alpha\beta}+(\tau_1)_{\alpha\beta}]
\sigma_\beta^H
\ee 
where  $\tau_1$ is the customary Pauli matrix, the superscript $1$ denotes the
order of correction, and
\be
\dl=\frac{g^2}{2}\int_M^E\frac{d^3\kvet}{2w_k(2\pi)^3}
\frac{2p_1p_2}{(kp_1)(kp_2)}=\frac{\alpha_W}{4\pi}\log^2\frac{E^2}{M^2}
\qquad (\alpha_W=\frac{g^2}{4\pi})
\ee
is the eikonal radiation factor for $W$ exchange.

From Eq. (\ref{firstorder}) we get in particular:
\be
\Delta\sigma_e^1= \dl (\sigma_\nu^H-\sigma_e^H)=-\Delta\sigma_\nu^1
\ee
which was the main result of \cite{3p}, leading also to
$\Delta\sigma_e^1+\Delta\sigma_\nu^1=0$, 
i.e., to cancellation upon isospin averaging.

The insertion mechanism just explained can be iterated to higher orders in the
strong ordering region $M\ll w_1\ll w_2.....\ll E$, by noticing that the
current (\ref{correnti}), after cancellation of the $\gamma,Z$ contributions
takes up only a $W$ part with only one cutoff ($M$), which cannot induce
symmetry breaking. Since the action of real $W$ emission is simply that of
interchanging $\nu$ and $e$ indices, we end up with a two channel problem
where the energy evolution hamiltonian - derived from eq. (\ref{firstorder})
 -  commutes at 
different energies. The outcome is 
a simple exponentiation of the first order result:
\be
\sigma_\alpha(s)=\sigma_\alpha^H+\sum_{n=0}^\infty\Delta\sigma_\alpha^n=
\{\exp\dl(\tau_1-1)\}_{\alpha\beta}\sigma_\beta^H\quad ,
\ee 
or, by simple algebra,
\be\label{eq:mp}
\sigma_{e,\nu}=\frac{\sigma_\nu^H+\sigma_e^H}{2}\mp
\frac{\sigma_\nu^H-\sigma_e^H}{2}e^{-2\dl}
\ee

This means that on average the double logs cancel out 
($\sigma_{e\bar{e}}+\sigma_{\nu\bar{e}}=
\sigma_{e\bar{e}}^H+\sigma_{\nu\bar{e}}^H$), while the
$\nu$-beam and $e$-beam difference $\sigma_{\nu\bar{e}}-\sigma_{e\bar{e}}$
decreases exponentially with the universal exponent $2\dl$, and vanishes
eventually at infinite energy.

Apart from the important phenomenological implications of Sec.6, the above
results have a nice theoretical interpretation in a non abelian framework,
which will be illustrated in the following sections. First, the 
exponent $2\dl$ can be related to the form factor in the adjoint 
representation, which is in fact a typically non abelian quantity; 
because of gauge invariance the exponent is also 
universal, i.e. the same for any fermion doublet in the initial state.
 A similar relationship was noticed for QCD
in \cite{mueller,ccm}. Furthermore, the
asymptotic equality of neutrino and electron beam cross sections is related to
the idea that in the unbroken limit ($M\to 0$), nonabelian charges 
are actually not observable as 
asymptotic states, as already noticed in QCD in
connection with preconfinement 
\cite{av} and factorization violating
contributions \cite{ccm}. In fact, if there were no symmetry breaking 
(infinite $2\dl$), Eq. (\ref{eq:mp}) would imply that any coherent 
superposition of electron and
neutrino states is projected by the nonabelian interaction into an incoherent
mixture with equal weights.

In the following sections we shall further elaborate on the heuristic result
of Eq.  (\ref{eq:mp}), by providing a more general classification, and a 
proof based on the coherent state operator formalism adopted for the
cancellation theorems of Sec.2.

\section{Leading coherent state operators}

The evaluation of BN violating terms - just pictured in the diagrammatic
approach - rests on the separation of initial and final state soft
interactions, that can be performed on the basis of the time evolution in a
given reference frame, e.g. the c.m. frame of the initial state. This method,
first applied by Faddeev and Kulish \cite{fk} to QED, has been extended to
(unbroken) non abelian theories in Refs. \cite{cm}, by constructing the large
time (asymptotic) Hamiltonian as an infinite series of progressively
subleading contributions.

Here we limit ourselves to describe such analysis at the leading (double log)
level. This does not mean  that subleading contributions are unimportant; it
just means that the state of the art in a broken theory is not sophisticated
enough, at present, to allow confidence in all-order subleading evaluations,
even for the inclusive quantities that we investigate here. In particular,
collinear factorization theorems should be carefully revised, partly because
of the very presence of the uncanceled double logs that we are computing. 

\subsection{Quantum Electrodynamics}
According to Eq. (\ref{eq:2}), the process is separated into a hard scattering
matrix $S_H$, involving the scale $E\gg\lambda$ ($\lambda$ is the IR cutoff),
and in a soft interaction, described by ${\cal U}^I$ and ${\cal U}^F$,
involving photon frequencies $w$ such that $E\gg w\gg \lambda$. In describing
the soft interaction, the hard (initial and final) fermions 
are treated as external currents. 
Thus, the
interaction Hamiltonian describing the evolution in the soft Hilbert space
${\cal H}_s$ is simply:
\ba\label{hamiltonian}
H_s(t)&=&\sum_ie_i\int_\lambda^Ed[k]\hat{p}_i^\mu\left(A_\mu(\kvet)
e^{-i(\hat{p}_ik)t}+h.c.\right)\equiv
\int d\nu(h_+(\nu)e^{-i\nu t}+h_-(\nu)e^{i\nu t})
\\
h_+(\nu)&=&\sum_ie_i\int_\lambda^Ed[k]\hat{p}_i^\mu
A_\mu(\kvet)\delta(\nu-\hat{p}_ik)
\ea
where $d[k]=\frac{d^3\kvet}{2w_k(2\pi)^3}$, 
$E_i\hat{p}_i^\mu=p_i^\mu$ ($i=1,...n_I$) denote the momenta of the hard
particles in the initial (or final) state, $k^\mu=(w_k,\kvet)$ is the photon
momentum, and $A_\mu(\kvet)$ ($A_\mu^\cro(\kvet)$) denote the photon
annihilation (creation) operators. The ``energy transfer''
variable $\nu=\hat{p}_ik$ 
is therefore conjugated
to the time variable $t$ in this approach, and 
represents physically the quantity of energy transferred in an elementary
vertex.

The key feature of Eq. (\ref{hamiltonian}) is the (universal) eikonal coupling
of charged hard particles to photons, proportional to their velocities 
$\hat{p}^\mu$. At amplitude level, this implies the insertion
formulas  with the eikonal current
\be J^\mu(k)=\sum_ie_i\frac{p_i^\mu}{kp_i}\quad ,\ee
whose non abelian counterpart has been given  in Eq. (\ref{correnti}).

The in (out) coherent state operators occurring in Eq. (\ref{eq:2}) 
are obtained
from the soft Hamiltonian (\ref{hamiltonian}) 
by computing the time-ordered evolution
operator before (after) the hard scattering. In the QED case, the calculation
is simplified by the fact that $H_s$ is linear in the $A_\mu$s, and has
therefore c-number commutators at non equal times. For instance we obtain, by
standard methods \cite{fk}, the initial state operator
\be
{\cal U}_I=U(0,-\infty)=e^{i\phi_C}
\exp\left[\int_\lambda^E\frac{d\nu}{\nu}(h_+(\nu)
-h_-(\nu))\right]\qquad\phi_C=2\pi\int_\lambda^E\frac{d\nu}{\nu}
[h_+(\nu),h_-(\nu)]
\ee

Let us now consider for simplicity the case of an initial state of two
particles 1 and 2 with opposite charges\footnote{For general charge
configurations, a gauge invariant picture requires considering initial and
final state operators all together}
and relative velocity $v_{12}$. 
By using the explicit form of the $h$'s we obtain:
\be\label{eq:3.4}
{\cal U}_I=U_1U_2=\exp\int_\lambda^Ed[k]J^\mu_{12}(k)
(A_\mu(\kvet)-A^\cro_\mu(\kvet))
=\,:\! {\cal U}_I\!:\,\exp[\int_\lambda^Ed[k][J_{12}(k)]^2
\ee
\ben
\phi_C=\frac{\alpha}{2v_{12}}\qquad J_{12}^\mu=
e(\frac{p_1^\mu}{kp_1}-\frac{p_2^\mu}{kp_2})
\een
Since the Coulomb phase $\phi_C$ doesn't give 
physical effects for
the processes considered here, we shall drop it from now on.

Note that the operator ${\cal U}_I$ is factorized in the particle indices
$i=1,2$ and is a functional of the 
combination $A_\mu(\kvet)-A_\mu^\cro(\kvet)\equiv i
\Pi_\mu(\kvet)$ only. For this reason 
the currents $J_i^\mu$ can be freely added
in the exponent, and unitarity is trivial. 

A less trivial, but straightforward step is the normal ordering quoted 
in Eq. (\ref{eq:3.4}), which emphasizes the Sudakov form factor. In the
massless limit we obtain
\be \label{Fa}
F_{12}\equiv\bra 0|U(1)U(2)|0\ket=\exp\left[
-e^2\int_\lambda^Ed[k]\frac{2p_1p_2}{(kp_1)(kp_2)} \right]
\approx\exp[-\frac{\alpha}{4\pi}\log^2\frac{E^2}{\lambda^2}]
\ee
From the same normal ordering formula one can obtain the (Poissonian)
properties of soft photon radiation, the energy resolution form factors, and
so on. 
\subsection{Non abelian coherent states}

The asymptotic Hamiltonian has been constructed \cite{cm}
in the (unbroken) SU(N)
gauge theory also. The difficulty, in such case, is that soft bosons can be
either primaries (i.e. emitted directly by the fast incoming particles) or
secondaries (i.e. emitted by the primary ones which have higher energy).
If strong ordering in energies is assumed, the soft Hamiltonian has
nevertheless a simple form, as follows:
\be\label{eq:24}
H_s(t)=g\int_\lambda^Ed[k]
\left[
\left(
\sum_it_i^a\hat{p}_i^\mu e^{-i(\hat{p}_ik)t}
+\int_{w_k}^Ed[k']\rho^a(\kvet')\hat{k}'^\mu e^{-i(\hat{k}'k)t}
\right)A_{\mu a}(\kvet)+\mbox{ h.c.}
\right]
\ee
\ben
\rho^a(\kvet)=-A_{\mu b}^\cro(\kvet)(T^a)^{bc}A_c^\mu(\kvet)\quad ,\quad
(T^a)^{bc}=if^{bac}
\een
In Eq. (\ref{eq:24}), 
the second term in round brackets represents the secondary
emission by primary bosons with energies $w_{k'}\gg w_k$, which occurs again
through an eikonal vertex $\hat{k}'^\mu\equiv\frac{k'^\mu}{w_{k'}}$.
Because of the  assumption of energy ordering, we are limiting ourselves to
the leading coherent state operators, and we refer to \cite{cm} for the
subleading contributions.

At this level of accuracy, we can compute the evolution operator 
by assuming
strong ordering in the energy transfers also (for details, see \cite{cm}). 
By indicating 
with $P_{\omega}$ the energy ordering operator (smaller energies act first), and by
introducing the field
$\Pi_{\mu a}(k)\equiv -i(A_{\mu a}(k)-A^\cro_{\mu a}(k))$, we obtain 
\ba\nonumber
U_s(0,-\infty) & = &
P_\nu \exp\left[\int_\lambda^E\frac{d\nu}{\nu}\left(h_+(\nu)
-h_-(\nu)\right)\right]\equiv V_s^EU_I^{l}
\\\label{eq:3.7}
&=&  P_{\omega}\exp 
\left[ig\int_\lambda^Ed[k]
\left(
\sum_it_i^a\frac{p_i^\mu}{kp_i}
+\int_{w}^Ed[k']\rho^a(\kvet')\frac{k'^\mu}{kk'}
\right)\Pi_{\mu a}(\kvet)
\right]\quad ;
\\\nonumber
V_s^E &\equiv &  P_{\omega}\exp 
\left[ig\int_\lambda^Ed[k]\int_{w_k}^Ed[k']
\rho^a(\kvet')\frac{k'^\mu}{kk'}
\Pi_{\mu a}(\kvet)
\right]
\ea

Here we face the problem of explicitating the recursive emission of
secondaries. We treat the full path-ordered exponential in Eq.~(\ref{eq:3.7})
as a two-potential problem, one of which produces the purely soft operator
$V_s$, which carries no isospin/color indices, and the other one is computed in
the interaction representation of the first, and provides the
properly called coherent state operator ${\cal U}_I^l$, carrying the 
isospin/color 
indices of the incoming particles. The result of this
procedure is
\ba\label{3.8}
{\cal U}_I^{l}&\equiv&V_s^\cro U_s(0,-\infty)=
P_{\omega}\exp\left[ig\int_\lambda^Ed[k]\sum_It_i^a\frac{p_i^\mu}{kp_i}
\Pi^{w_k}_{\mu a}(\kvet)\right]
\\
{\cal U}_F^{l}&\equiv&U_s^\cro(0,-\infty)V_s=
\bar{P}_{\omega}\exp\left[-ig\int_\lambda^Ed[k]\sum_Ft_f^a\frac{p_f^\mu}{kp_f}
\Pi^{w_k}_{\mu a}(\kvet)\right]
\ea
where $\bar{P}_{\omega}$  is the energy anti-ordering operator 
(bigger energies act first) and where the ``dressed'' field
$\Pi^{w}_{\mu a}$
can be made more explicit as follows
\be\label{eq:3.9}
\Pi^{w}_{\mu a}\equiv V_s^{w\cro}\Pi_{\mu a}V_s^{w}
=({U}^{\omega}_A)_{ab}(\hat{\kvet})\Pi_{\mu b}(\kvet)
\ee
where $({U}^{\omega}_A)_{ab}(\hat{\kvet})$ denotes the coherent state operator
in the {\sl adjoint} representation, for a boson $w\hat{k}^\mu$, having the
matrix properties (recall $(T_a)_{bc}=if_{bac}$)
\be{ U}_{A}^*={ U}_{A}\qquad{U}_{A}^T
={U}_{A}^\cro={U}_{A}^{-1}\label{3.10}\ee
Eq. (\ref{eq:3.9}) was proved in Ref \cite{cm}, by showing that the 
l.h.s. and the r.h.s. satisfy the same evolution equation 
in the $w$ variable.

Eqs. (\ref{3.8}-\ref{3.10}) show that non abelian coherent states operators
are defined in a quite nonlinear manner. Eq. (\ref{3.8}) exponentiates up to
energy $E$ the dressed fields $\Pi^{\omega}$, which in turn are expanded in
terms of a coherent state at a lower energy $w<E$ by Eq. (\ref{eq:3.9}), which
in turn etc......
In other words non abelian soft radiation is described 
by a ``coherent state of lower energy coherent states''
whose structure is obtained recursively by a branching process \cite{cm}, 
by now incorporated in some QCD event generators \cite{MW}.

For our purpose here, the important point is that we are now able to
explicitate the soft part of the process (\ref{eq:1}) as follows
\be
S=U_s^\cro(0,+\infty)S_HU_s(0,-\infty)={\cal U}_F^lS_H{\cal U}_I^l
\ee
Here the operator $V_s$ defined in Eq. (\ref{eq:3.7}),
being purely soft, commutes with $S_H$ and
drops out by unitarity, and ${\cal U}_F^l,{\cal U}_I^l$ are provided by 
Eqs. (\ref{3.8}) and (\ref{eq:3.9}) at leading level. 

The leading operator ${\cal U}^l$ - despite its nonlinearity - is still a
functional of the fields $\Pi_{\mu a}(\kvet)$ only, and involves therefore
only {\sl commuting} quantities in the Fock space. It follows, therefore, that
commutativity for any given color indices holds:
\be
[{\cal U}^l_{\beta\alpha},{\cal U}^l_{\beta'\alpha'}]=0
\ee 
as already used in Sec.2, and that factorization with respect to the hard
particles holds also
\be
{\cal U}^l_{\beta\alpha}(1,\dots,n)={\cal U}^l_{\beta_1\alpha_1}(1)
\, {\cal U}^l_{\beta_2\alpha_2}(2)\,\dots\,
{\cal U}^l_{\beta_n\alpha_n}(n)\qquad .
\ee

Finally, one should note that the coherent state in the adjoint 
representation regulates the evolution equation of $t^a$ matrices:
\be\label{3.14}
{U}^{E\cro}(\pvet)t_a{U}^{E}(\pvet)
=({U}^E_A)_{ab}(\pvet)t_b
\ee
for an arbitrary isospin/color representation of the particle $\pvet$. 
In fact, the coherent state operators satisfy the Schr\"{o}dinger-like
equation
\be\label{3.15}
\frac{\de}{\de E}U^E(\pvet)=t_b\Delta_b^EU^E(\pvet)
\qquad
\Delta_b^E=ig\int d[k]\delta(w_k-E)\frac{p^\mu}{kp}\Pi_{\mu b}^E(\kvet)
\ee 
Therefore, by combining the $U$ and $U^\cro$ equations, the l.h.s. of
Eq. (\ref{3.14}) satisfies the equation
\be
\frac{\de}{\de E}\mbox{Tr}(U^{\cro E}t_aU^Et_d)=
if_{abc}\Delta_b^E\mbox{Tr}(U^{\cro E}t_cU^Et_d)
\ee 
which is the same as the one satisfied by the r.h.s., by direct use of
(\ref{3.15}) in the adjoint representation.

\subsection{Form factor exponentiation}
The Sudakov (singlet) form factor can be defined similarly to Eq. (\ref{Fa})
(with opposite initial charges), for any given color representation of
particles 1 and 2:
\be\label{36}
\bra 0 |\left(U^{E\cro}(2)U^E(1)\right)_{\alpha\beta}
|0\ket=\delta_{\alpha\beta}F_{12}(E,\lambda)
\ee
However, due to the nonlinearity of Eqs. (\ref{3.8}) and
(\ref{eq:3.9}) the normal ordering is no longer straightforward, and was
worked out by an evolution equation method in Ref. \cite{cm}.

Since the energy ordered exponential satisfies the Schr\"odinger-like
equation (\ref{3.15}), one first computes the energy derivative of 
(\ref{3.14}), using also definition (\ref{eq:3.9}), as follows ($w_K=E$)
\be
\mbox{Tr}(1)\frac{\de}{E\de E} F_{12}
=\bra 0 |\mbox{Tr}\left(U^{E\cro}(2)\frac{gd\Omega_K}{2(2\pi)^3}
J_{12}^\mu t_a (U_A)_{ab}^E(\hat{K}) 
(A_{\mu b}(\Kvet)-A^\cro_{\mu b}(\Kvet))U^E(1)
\right)|0\ket 
\ee
where  $J_{12}^\mu = \frac{p_1^\mu}{Kp_1} - \frac{p_2^\mu}{Kp_2}$. 
Then, one commutes $A_{\mu b}(K)$ to the right and $A^\cro$ to the left. Since
$[A_{\mu b}(\Kvet),A^{w_k\cro}_{\nu a}(\kvet)]=0$ 
for $w_k<E$ this procedure singles out the upper frequency in
$U^E$ and yields
\be
[A_{\mu b}(\Kvet), U^E(1)]=g(U_A)_{cb}^E
\frac{p_1^\mu}{Kp_1}t_c U^E(1)
\ee
with a similar relation involving $A^\cro$ and $p_2^\mu$.
Finally, by using the unitarity relation
\be\label{3.20}
\sum_b (U_A^E)_{ab}(i)(U_A^E)_{cb}(i)=\delta_{ac}
\qquad
(i=1,2)
\ee
and $\tvet^2_i=C_F$ (or $C_A$, depending on the particle's representation) 
we obtain
\be\label{3.21}
\frac{\de}{E\de E}F_{12}(E,\lambda)=\frac{g^2 C_{F(A)}}{2(2\pi)^3}
\left(\int_{w_K=E} d\Omega_K(J_{12}(K))^2\right)F_{12}(E,\lambda)
\ee
which is the evolution equation for the form factor we were looking for. In
the double log approximation, (\ref{3.21}) yields
\be\label{41} F_{12}(E,\lambda)=
\exp[-\frac{g^2C_{F(A)}}{16\pi^2}\log^2\frac{E^2}{\lambda^2}]
\ee
as expected. Note that, in the $SU(2)$ isospin case, $C_A=2$ and the exponent
in Eq. (\ref{41}) becomes just $2\dl$. 

The essential point of this derivation rests on the unitarity relation
(\ref{3.20}), which means the cancellation of all correlation effects due to
the nonabelian structure. This leads to Eq. (\ref{3.21}), which could be
naively derived by the ``external line insertion rule'' 
for virtual corrections.
\section{Bloch Nordsieck violation to all orders}

\subsection{Isospin structure of the hard overlap matrix}
In this section we discuss the general structure of the hard 
overlap matrix, describing the squared matrix element:
$O_H\equiv S_H^\cro S_H$. We consider the case  of   a generic
process with two partons in the initial state and  
we work in a limit in which all kinematical
invariants are much bigger than gauge bosons
masses: $|s|,|t|,|u|\gg M^2$. Then, the full gauge symmetry
SU(3)$\otimes$SU(2)$\otimes$U(1) is restored, and the structure of the overlap
matrix in isospin space is fixed by the SU(2) symmetry.

\begin{figure}[htb]\setlength{\unitlength}{1cm}
\begin{picture}(14,7.5)
\put(-.7,1){\epsfig{file=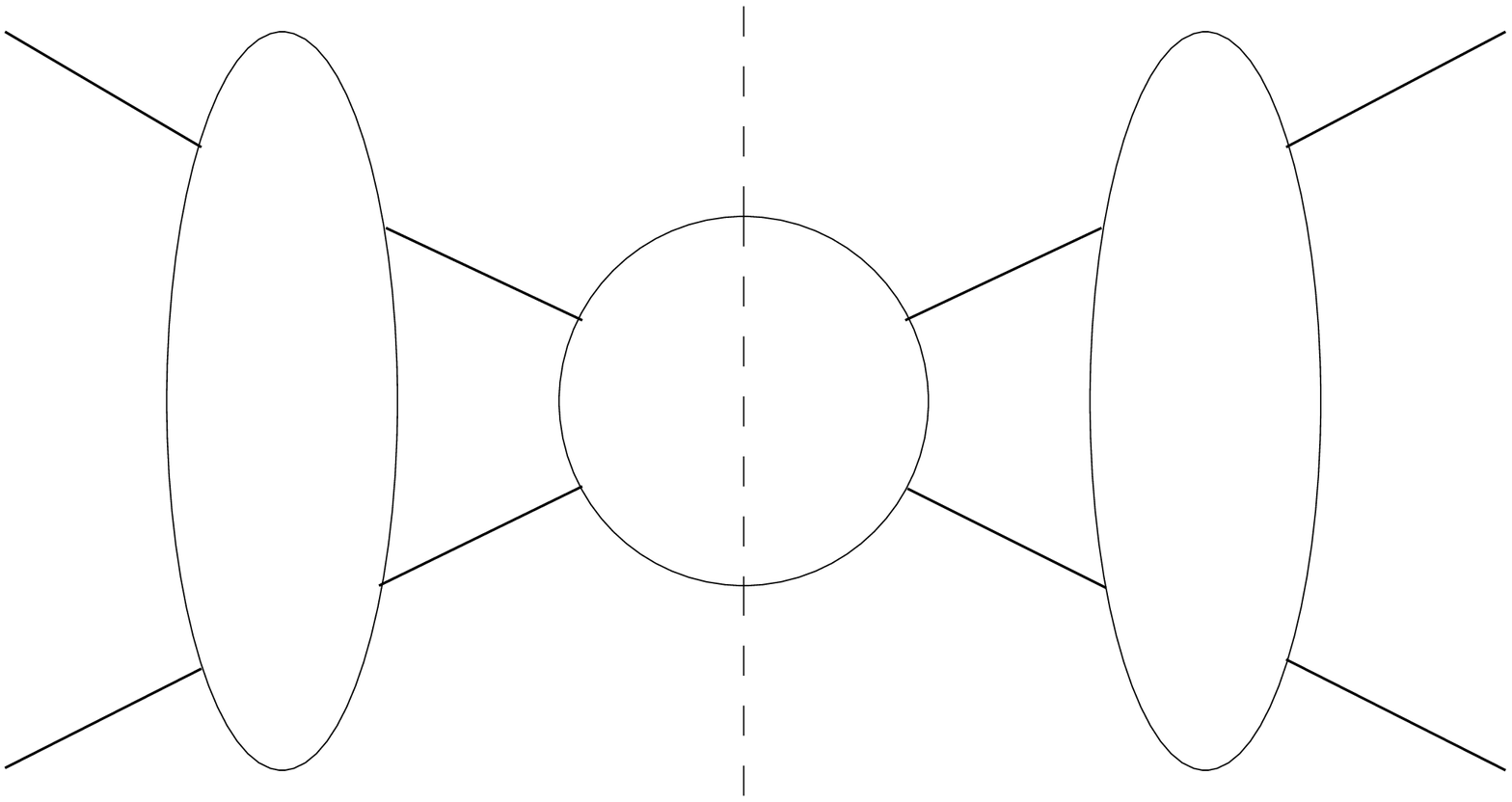,height=4.8cm}}
\put(-.8,5.7){$\alpha_1$}\put(-.8,1){$\alpha_2$}
\put(2,4.4){$\alpha'_1$}\put(2,2){$\alpha'_2$}
\put(5,4.4){$\beta'_1$}\put(5,2){$\beta'_2$}
\put(7.4,5.7){$\beta_1$}\put(7.4,1){$\beta_2$}
\put(.8,3.3){${\cal U}^I$}
\put(6,3,3){${\cal U}^{I\cro}$}
\put(3.1,3.3){$S_H^\cro\,\, S_H$}
\put(9,1){\epsfig{file=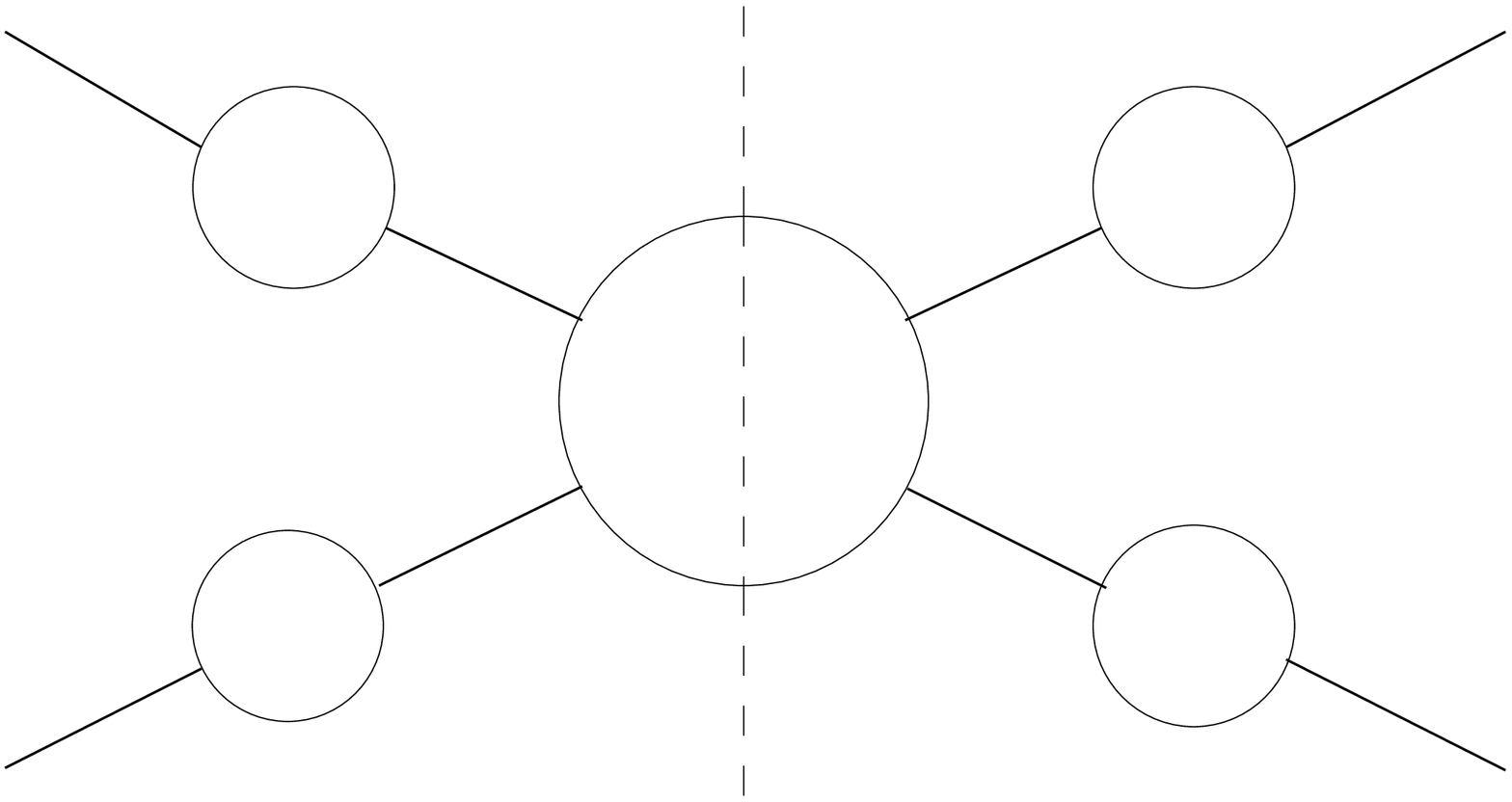,height=4.8cm}}
\put(12.8,3.3){$S_H^\cro\,\, S_H$}
\put(10.2,2){$U^{(2)\cro}$}\put(10.3,4.6){$U^{(1)}$}
\put(15.6,2){$U^{(2)}$}\put(15.5,4.6){$U^{(1)\cro}$}
\put(3.4,0.52){(a)}\put(13.1,.52){(b)}
\end{picture}
\caption{\label{figsoftdress}
Soft dressing of the hard S-matrix  $S_H$ 
is described by the coherent state 
operator ${\cal U}^I$ (a). At leading order, the latter is factorized
into  leg operators $U^{(i)}$ (b).}
\end{figure}

\begin{figure}[htb]\setlength{\unitlength}{1cm}
\begin{picture}(12,7.5)
\put(0,0){\epsfig{file=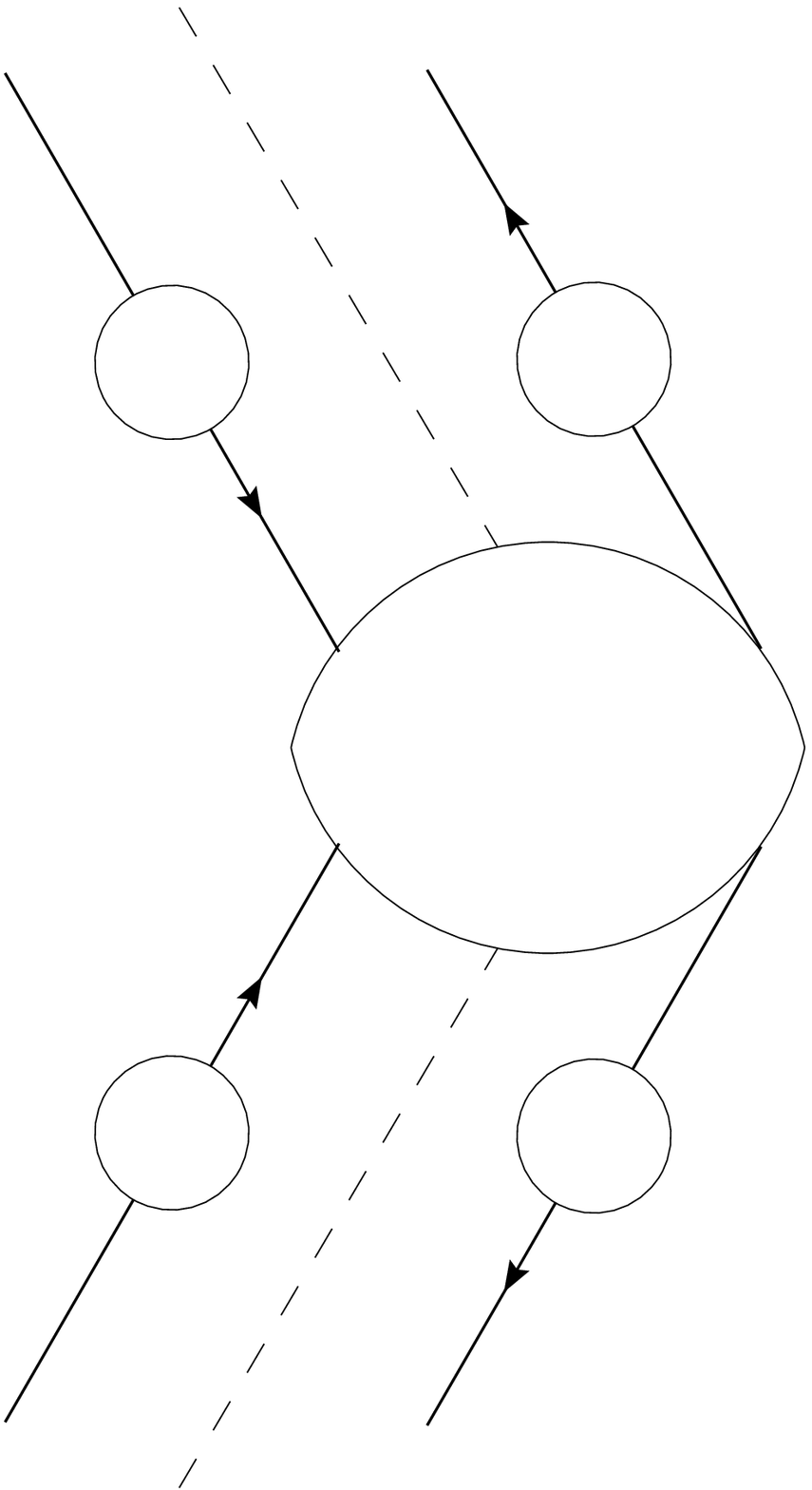,height=8cm}}
\put(.7,5.9){$U^{1}$}\put(2.9,5.9){$U^{1\cro}$}
\put(.7,1.8){$U^{2}$}\put(2.9,1.8){$U^{2\cro}$}
\put(2,3.9){$O_H\equiv S_H^\cro S_H$}
\put(6,3.9){{\huge $=$}}\put(12,3.9){{\huge $+$}}
\put(7,0){\epsfig{file=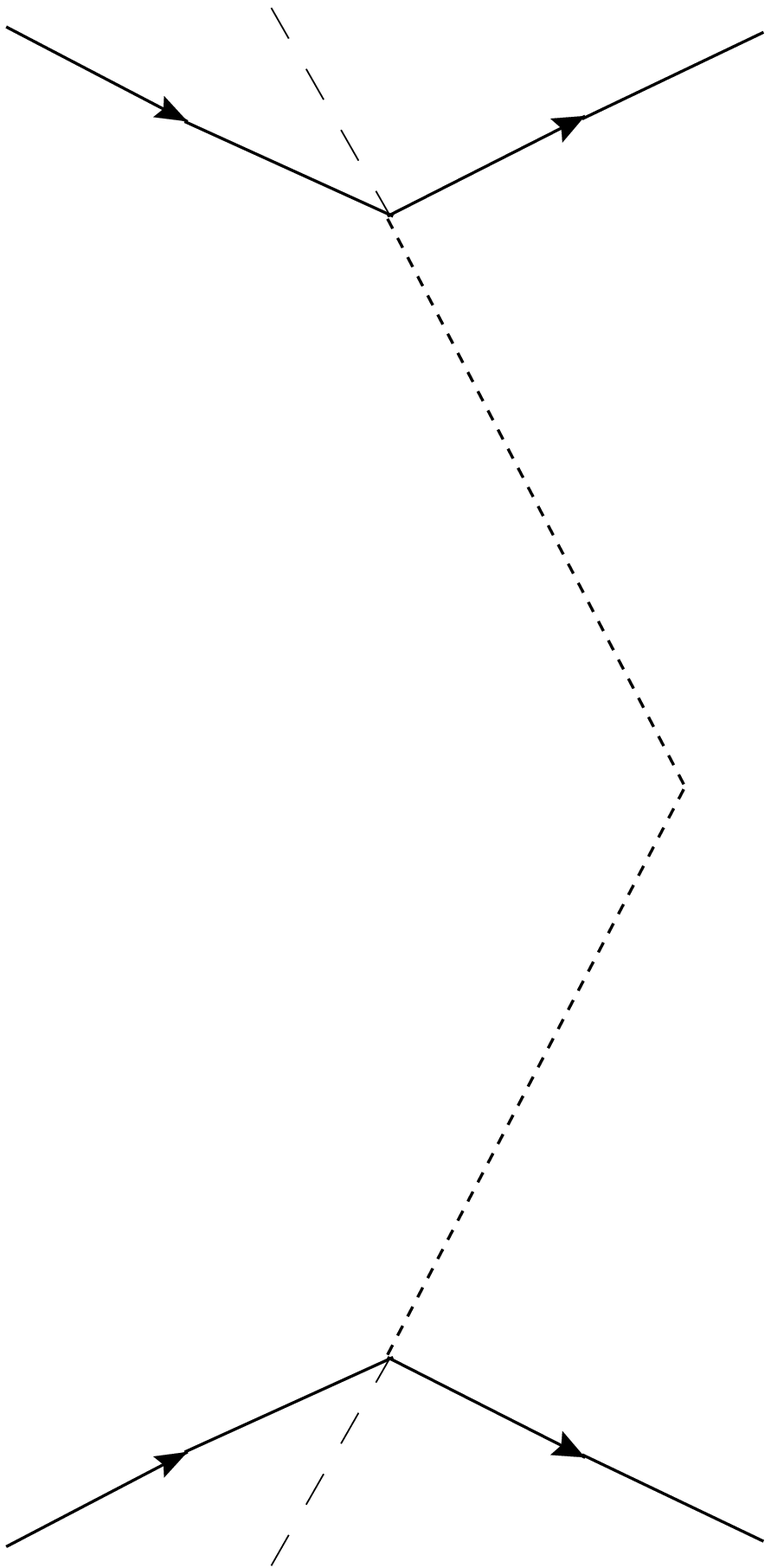,height=8cm}}
\put(13,0){\epsfig{file=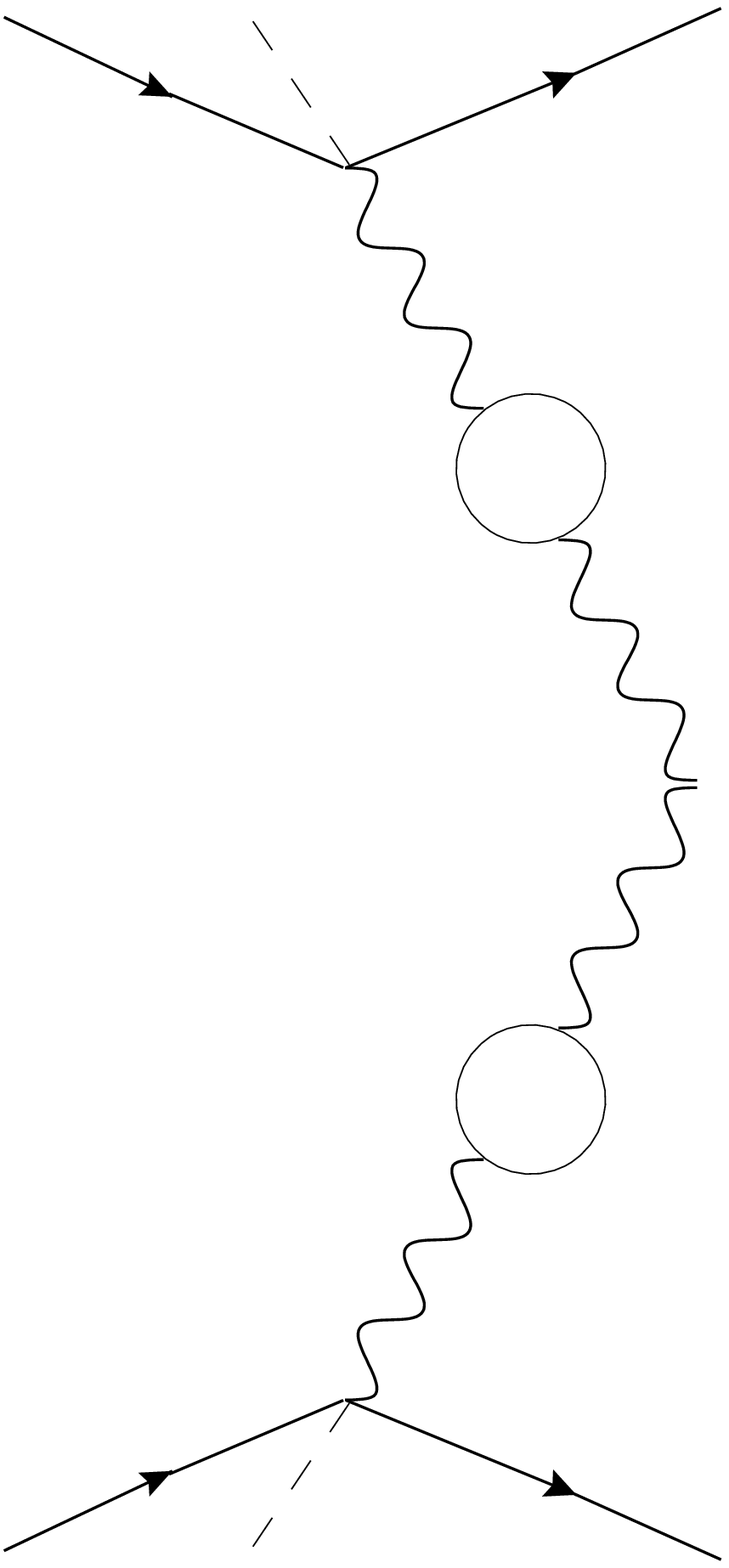,height=8cm}}
\put(15.4,5.5){$U_A^{1}$}\put(15.4,2.3){$U_A^{2}$}
\end{picture}
\caption{\label{figsinvec}
Singlet and vector decomposition of the overlap matrix $O_H$ in the
t-channel. The adjoint coherent state dressing is depicted also}
\end{figure}

Left particles carry nonabelian $SU(2)$ charges while right particles
don't, so we need to consider three cases:
\begin{itemize}
\item
when both initial particles are
righthanded, and therefore do not carry any nonabelian weak charge nor any
isospin index, the overlap matrix is simply a number $O^H$ in
isospin space (still depending of course on the quantum numbers of the 
involved particles). 
\item
Next possibility is that one particle is left polarized and the other
one is right polarized. In this case the hard overlap matrix carries two 
(left) isospin indices $O^H_{\beta\alpha}$.  
\item
The case of two left initial fermions, is of course the most complicated one;
the hard overlap matrix carries in this case four isospin indices
$O^H_{\beta_1\beta_2,\alpha_1\alpha_2}$ (see Fig \ref{figsoftdress}).
\end{itemize}
While for cross sections we always have $\alpha_i=\beta_i$, we leave open the
possibility that $\alpha_i\neq \beta_i$, and see $O^H$ as an operator in
isospin space. The form of the overlap matrix is severely restricted by the
requirement of SU(2) symmetry. In the three cases discussed above, we have:
\begin{equation}\label{decomp}
RR:O^H=A_0\qquad
RL,LR:O^H_{\beta,\alpha}=B_0\delta_{\beta\alpha}
\qquad
LL:O^H_{\beta_1\beta_2,\alpha_1\alpha_2}=C_0\delta_{\beta_1\alpha_1}
\delta_{\beta_2\alpha_2}+C_14t^a_{\beta_2\alpha_2}t^a_{\beta_1\alpha_1}
\end{equation}
The last expression, where the indices $\alpha_1,\beta_1$ (and
$\alpha_2,\beta_2$) are grouped together, corresponds to a t-channel
decomposition in singlet and vector components (see Fig. \ref{figsinvec}).
The case where the initial particle on leg 2 is an antiparticle, thus
belonging to the conjugate representation $t^*=t^T$,
is correspondingly:
\be\label{ultima}
\bar{O}^H_{\beta_1\beta_2,\alpha_1\alpha_2}=
\bar{C}_0\delta_{\beta_1\alpha_1}\delta_{\beta_2\alpha_2}
+\bar{C}_1\;4t^a_{\alpha_2\beta_2}t^a_{\beta_1\alpha_1}
\quad\qquad\mbox{ (part - antipart)}
\ee
Let us  consider as an example a generic hard cross section
involving a left $e^-$ and a left $e^+$ 
(which we indicate with $e$ and $\bar{e}$),
and $\nu,\bar{\nu}$\footnote{In our notation a particle and its antiparticle
share the same isospin index: 1 for $\nu,\bar{\nu}$ and 2 for $e,\bar{e}$}. 
We have:
\begin{subequations} \label{43}
\begin{eqalignno}
\sigma_{e\bar{e}}^H=\sigma_{\nu\bar{\nu}}^H\propto
\bar{O}^H_{11,11}=\bar{O}^H_{22,22}&=\bar{C}_0+\bar{C}_1
\\
\sigma_{e\bar{\nu}}^H=\sigma_{\nu\bar{e}}^H\propto
\bar{O}^H_{12,12}=\bar{O}^H_{21,21}&=\bar{C}_0-\bar{C}_1
\end{eqalignno}
\end{subequations}

\subsection{Resummed energy dependence from coherent states}
\begin{figure}[t]\setlength{\unitlength}{1cm}
\begin{picture}(12,6)
\put(3.5,.5){\epsfig{file=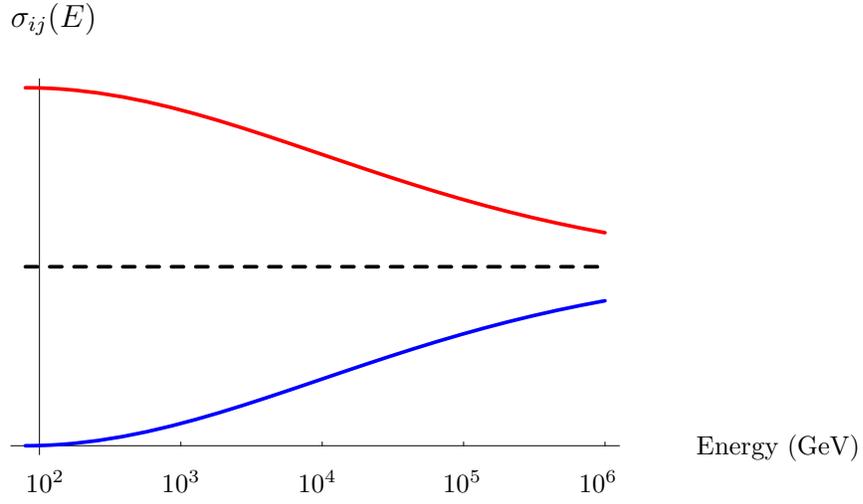,height=5cm}}
\put(3.5,6.2){\large $\sigma_{ij}(E)$}
\put(3.7,0){$10^2$\hskip1.3cm$10^3$\hskip1.3cm$10^4$
\hskip1.3cm$10^5$\hskip1.3cm$10^6$ }
\put(12.5,0.5){ Energy (GeV) }
\end{picture}
\caption{\label{sigma12}$\sigma_{12}$ and $\sigma_{11}$ as a function of
energy. The vertical scale is arbitrary}
\end{figure}

We now proceed to ``dress'' the hard matrix element with soft 
interactions.
In the case of right initial particles, 
since weak interactions become purely abelian in this case, it
should be clear from the above discussion that no BN-violating effect is
present. 
If one particle is L and the other one is R, 
the dressing by soft interactions is
described by (\ref{eq:5}):
\be\label{44}
O^H_{\alpha\beta}\stackrel{dress}{\to}
\;O_{\alpha\beta}=\;_S\bra 0|
{\cal{U}}^{\cro}_{\alpha\alpha'}O^H_{\alpha'\beta'}
{\cal{U}}_{\beta'\beta}|0\ket_S
\ee
But then, since by SU(2) symmetry
$O^H_{\alpha\beta}=B_0\; \delta_{\alpha\beta}$,
and  because of the unitarity property
(\ref{eq:3}), also in this case no BN violating effect is present and the
dressed overlap matrix is equal to the hard one. In the remaining of this
section we discuss the interesting case of two left initial fermions.
As we have seen, the dressing in this case
is described by a coherent state operator 
${\cal U}^I$ such that (see Fig. \ref{figsoftdress}):
\be\label{dress}
O^H_{\beta_1\beta_2,\alpha_1\alpha_2}
\stackrel{dress}{\rightarrow}O_{\beta_1\beta_2,\alpha_1\alpha_2}=
_S\!\bra 0|{\cal U}^{I\cro}_{\beta_1\beta_2,\beta_1',\beta_2'}
(O_H)_{\beta_1'\beta_2',\alpha_1'\alpha_2'}
{\cal U}^I_{\alpha_1'\alpha_2',\alpha_1,\alpha_2}|0\ket_S
\ee
where $|0\ket_S$ is the soft vacuum. 
At the leading log level, ${\cal U}^{I\cro}$ is factorized (Sec. 4.2)
into two leg operators:
\be\label{penultima}
{\cal U}^I_{\alpha_1'\alpha_2',\alpha_1,\alpha_2}=
U^{(1)}_{\alpha_1'\alpha_1}U^{(2)\cro}_{\alpha_2\alpha_2'}\quad
\mbox{part-antipart}\qquad
{\cal U}^I_{\alpha_1'\alpha_2',\alpha_1,\alpha_2}=
U^{(1)}_{\alpha_1'\alpha_1}U^{(2)}_{\alpha_2'\alpha_2}\quad
\mbox{part-part}
\ee
where we take into account that antiparticles live in the conjugate
representation, so that in the particle-antiparticle case
we have $U^{(2)}\to U^{(2)*}$.
Putting together (\ref{ultima},\ref{dress},\ref{penultima}) 
we obtain, for the part -
antipart case,  the dressed overlap operator:
\begin{equation}
           \begin{eqalign}
          \bar{O}_{\beta_1\beta_2,\alpha_1\alpha_2} &\equiv
\bar{{ C}}_0(s)\delta_{\beta_1\alpha_1}\delta_{\beta_2\alpha_2}
+\bar{{C}}_1(s)\;4t^a_{\beta_1\alpha_1}t^a_{\alpha_2\beta_2}
\\
  &= \bar{C}_0\delta_{\alpha_1\beta_1}\delta_{\alpha_2\beta_2}
+4 \bar{C}_1\;_S\!\bra 0|(U^{(1)\cro}\, t^a \, U^{(1)})_{\beta_1\alpha_1}
(U^{(2)\cro} \,t^a\,U^{(2)})_{\alpha_2 \beta_2 }|0\ket_S
           \end{eqalign}
\end{equation}
By using twice Eq. (\ref{3.14}) that relates the coherent states in the
fundamental representation with the one in the adjoint representation, we
obtain: 
\be
_S\!\bra 0|(U^{(1)\cro} t^a  U^{(1)})_{\beta_1\alpha_1}
(U^{(2)\cro} t^a\,U^{(2)})_{\alpha_2\beta_2}|0\ket_S=
_S\!\bra 0|\left(U_A^{(2)\cro}U_A^{(1)}\right)_{ab}|0\ket_S
t^a_{\beta_1\alpha_1}t^b_{\alpha_2 \beta_2}=
F_A(s,M^2)t^a_{\beta_1\alpha_1}t^a_{\alpha_2 \beta_2 }
\ee
where the definition (\ref{36}) of the form factor has been used, so that we
obtain:
\be\label{49}
\bar{C}_0(s)=\bar{C}_0\qquad\qquad \bar{C}_1(s)
=\bar{C}_1 e^{-C_A\dl}=\bar{C}_1 e^{-2\dl}
\ee
From these expression we obtain the final results for 
the dressed cross sections (see Fig. \ref{sigma12}):
\begin{subequations} \label{risfin}
\begin{eqalignno}
\sigma_{11} =\sigma_{22} & =  
\bar{C}_0(s)+\bar{C}_1(s)=\bar{C}_0+\bar{C}_1  e^{-2\dl}
=\frac{(\sigma_{11}+\sigma_{12})^H}{2}
+\frac{(\sigma_{11}-\sigma_{12})^H}{2} \; e^{-2\dl}
\\
\sigma_{12}=\sigma_{21} &=
\bar{C}_0(s)-\bar{C}_1(s)=\bar{C}_0-\bar{C}_1  e^{-2\dl}
=\frac{(\sigma_{11}+\sigma_{12})^H}{2}
-\frac{(\sigma_{11}-\sigma_{12})^H}{2} \; e^{-2\dl}
\end{eqalignno}
\end{subequations}
which reproduce Eq.(\ref{eq:mp}), and the relative effects in double log 
approximation:
\begin{subequations}  \label{risfin2}
\begin{eqalignno}
\left(\frac{\Delta\sigma}{\sigma}\right)_{11} & \equiv  
\frac{\sigma_{11}-\sigma_{11}^H}{\sigma_{11}^H}
=\left(\frac{\sigma_{11}^H-\sigma_{12}^H}{\sigma_{11}^H}\right)
\left(\frac{1-e^{-2\dl}}{2}\right)
\\
\left(\frac{\Delta\sigma}{\sigma}\right)_{12} & \equiv  
\frac{\sigma_{12}-\sigma_{12}^H}{\sigma_{12}^H}
=\left(\frac{\sigma_{12}^H-\sigma_{11}^H}{\sigma_{12}^H}\right)
\left(\frac{1-e^{-2\dl}}{2}\right)
\end{eqalignno}
\end{subequations}
An analogous treatment for the part - part case allows one to conclude that
Eqns. (\ref{risfin},\ref{risfin2}) 
hold also for  this case with the obvious replacement
$\bar{C}_i\to C_i$. These final results are therefore 
completely general: $\sigma_{11}$ stands for any cross
section with two incoming particles (or one particle and one antiparticle)
with isospin index 1, which means for instance $\sigma_{\nu\bar{\nu}},
\sigma_{uu},\sigma_{u\bar{u}}$ and so on. While in all these cases the
expressions for the coefficients 
$C_0$ and $C_1$ (or $\bar{C_0}$ and $\bar{C_1}$) that describe the hard cross
section are of course
different in general, the expression for the dressed cross sections 
is always the same and is described by  Eqns.  (\ref{risfin}).

Notice the appearance, as anticipated, of the adjoint Casimir $C_A=2$
in (\ref{risfin}). This means that the energy dependence of the effect we are
discussing is {\it universal}, i.e. the same for any fermion doublet in the
initial state. Note however that the relative effect does depend on the
structure of the hard cross sections, as one can see from (\ref{risfin2}). We
will discuss several cases of phenomenological interest
more in detail in Sec. 6.

\section{Applications to simple processes}
We consider inclusive observables associated  
to a large angle hard scattering process ($|s|\sim |t|\sim |u|\gg M^2$)
involving massless 
fermions and antifermions in the initial and final states. As we
have seen, a BN violating effect is present only if there are {\it two}
nonabelian charges in the initial state. This means that big noncancellations
are present only with two left fermions in the initial state, while the effect
is absent in the RR and RL cases. In turn, this implies a strong dependence on
the physical polarization of the initial beams: a maximal effect if the beams
are both polarized L, no effect in all other cases, and somewhere in between
for unpolarized beams. This is particularly important since one of 
NLCs features
is the possibility of having highly polarized beams \cite{NLC}.

We discuss in this section some cases that we think
are/will be phenomenologically relevant for NLCs (Sec. 6.1), 
and for electron-hadron (Sec. 6.2) and hadron-hadron (Sec. 6.3) colliders.
Given the master formulas (\ref{risfin}), and since the energy dependence is
universal as already noticed,
it is clear that only the (tree level) hard cross sections need to be
discussed. We expect large effects when LL contributions dominate 
the hard cross section, and/or when  
there is a big difference between $\sigma_{12}$ and $\sigma_{11}$ 
(see (\ref{risfin2})). The latter is the case, for instance, in pure 
$q\bar{q}$ s-channel annihilation where $\sigma_{11}$ is of order $\alpha_S^2$
while $\sigma_{12}$ is flavor changing and  
is thus electroweak (Sec. 6.3). Similarly,
in  $e^+e^-\to$ hadrons (Sec. 6.1), the effect is pretty large because this
process is dominated by L components.

\subsection{$l\bar{l}\to q\bar{q}$ (s-channel) annihilation}
This kind of process is typically relevant for NLC.  The Born (hard) 
amplitude is simply
described by an s-channel annihilation involving only weak interaction 
and has the form (a=1,2,3):
\be\label{6.1}
{\cal M}_{\beta_1\beta_2,\alpha_1\alpha_2}\sim
g'^2yY\delta_{\beta_1\beta_2}\delta_{\alpha_1\alpha_2}
+g^2t^a_{\beta_1\beta_2}t^a_{\alpha_1\alpha_2}
\ee
where $y$ ($Y$) denote the initial lepton (final quark) hypercharges. 
Expression (\ref{6.1}) is the s-channel analogue of
the singlet-vector t-channel decomposition (\ref{decomp}). By squaring
and summing over final states, 
we obtain the overlap matrix t-channel components, as defined in
(\ref{decomp}), for the case of (initial) L fermions:
\begin{subequations}\label{55}
\begin{eqaligntwo}\label{schannel}
C_1^{L} &=-\frac{1}{16}{g^4}+\frac{1}{2}{g'^4}y_L^2\sum Y_L^2 
&
C_0^{L} &=\frac{3}{16}g^4+\frac{1}{2}g'^4y_L^2\sum Y_L^2 \\
C_0^{R}&= \frac{1}{2} g'^4y_L^2 \sum Y_R^2 \qquad
&
C_1^{R}&=  \frac{1}{2} g'^4y_L^2 \sum Y_R^2
\end{eqaligntwo}
where $C_i^{L(R)}$ refers to final L(R) quarks.
For the case of initial R leptons we obtain:
\be\label{schannel-c}
A_0^{R}=\delta_{ij}\, g'^4y_R^2\sum Y_R^2\qquad\qquad
A_0^{L}=\delta_{ij}\, g'^4y_R^2\sum Y_L^2
\ee
\end{subequations}
where the factor $\delta_{ij}$ in (\ref{schannel-c}) takes into account that
the contribution $\propto g'^4$ is present only for the case of an initial
particle and its own antiparticle. The sums are over final states
hypercharges, namely $\sum Y_R^2=\frac{4}{9}+\frac{1}{9}$ and
$\sum Y_L^2=\frac{1}{36}+\frac{1}{36}$ for quarks. 
The unpolarized cross section is then found by putting in the usual phase
space factor, and by using the resummed energy
evolution in Eq. (\ref{49}). We find:
\begin{equation}\label{6.5}
\frac{d\sigma_{ij}}{d\cos\theta}=\frac{N_cN_f}{128\pi s}
\left[
(A_0^{R}+C_0^{L}\pm C_1^{L}e^{-2\dl})
(1+\cos\theta)^2
+(A_0^{L}+C_0^{R}\pm C_1^{R}e^{-2\dl})
(1-\cos\theta)^2\right]
\end{equation}
where $N_f$ is the number of quark families, $N_c$ is the number 
of colors and the + (-) sign
refers to $\sigma_{11}$ ($\sigma_{12}$).

The terms proportional to $g'^4$ in Eq. (\ref{6.5})
are pretty small, being suppressed by a $\frac{g'^4}{g^4}=t_W^4$ 
factor. In the limit 
$g'\to 0$ only L components contribute to the cross section:
\be\label{simplified}
\frac{d\sigma_{ij}}{d\cos\theta}\approx
\frac{d\sigma_{ij}^{LL}}{d\cos\theta}=
\frac{\pi N_c N_f\alpha_W^2}{32 s}\left(\frac{1+\cos\theta}{2}\right)^2
\left(3\mp e^{-2\dl}\right)
\ee
Note that, for instance,
$d\sigma_{\nu\bar{e}}^H>d\sigma_{e\bar{e}}^H$. Therefore, the BN
violating corrections {\it increase} the physically relevant 
$e\bar{e}$ cross section, that reaches in the asymptotic limit of very high
energy the isospin average. From (\ref{simplified}) we also 
see that in the $g'\to 0$ limit, angular and energy
dependences are factorized. Therefore in this limit the forward-backward
asymmetry for $e\bar{e}$ is equal to the tree level value of $\frac{3}{4}$.
However, the $g'^4$ terms are not completely negligible, producing a relative
correction to $A_{FB}$ of about 1.8 \% at 1 TeV, the dependence on
energy  being given by the
by now familiar universal behavior.

From (\ref{6.5}) and taking into account the 
$g'^4$ terms also, we obtain the relative effect for for 
$e_L\bar{e}_L\to$ hadrons:
\be\label{>?}
(\frac{\Delta\sigma_{e\bar{e}}}{\sigma^H_{e\bar{e}}})^{L}=
(\frac{\sigma_{\nu\bar{e}}^H-
\sigma_{e\bar{e}}^H}{\sigma_{e\bar{e}}^H})^{L }
\; (\frac{1-e^{-2\dl}}{2})
\approx 0.8\dl
\ee 
The effect for the unpolarized cross section is slightly reduced (we give also
the first order QCD corrections):
\be
\left(\frac{\Delta \sigma}
{\sigma}\right)^{EW}_{e\bar{e}}\simeq 0.58\;L_W(s)=
0.58\,\frac{\alpha_W}{4\pi}\log^2\frac{s}{M^2}\qquad\qquad
\left(\frac{\Delta \sigma}
{\sigma}\right)^{QCD}_{e\bar{e}}\simeq \frac{\alpha_S}
{\pi}
\ee
That is, radiative  corrections to $e^+e^-\to$ hadrons 
of {\it weak} origin  are bigger, at the TeV scale, than {\it strong} QCD
corrections (see Fig. \ref{had})

\subsection{t - channel scattering}
Here we consider processes in which the Born term is a
t-channel scattering diagram involving only weak interactions. 
This involves processes like $lq$ and $l\bar{q}$ scattering where $l$ is a
lepton and $q$ a quark. We also discuss briefly the case for 
$\nu_\mu e$ scattering.
The hard cross section is described in this case by the amplitude
\be\label{6.8}
{\cal M}_{\beta_1\beta_2,\alpha_1\alpha_2}\sim
g'^2yY\delta_{\beta_1\alpha_1}\delta_{\beta_2\alpha_2}
+g^2t^a_{\beta_1\alpha_1}t^a_{\beta_2\alpha_2}
\ee
where $y,\alpha_1$ ($Y,\alpha_2$) refer to lepton (quark)
indices. 
By squaring and by summing over final states 
we obtain the projections of the
overlap matrix for this case:
 \begin{subequations} \label{62}
       \begin{eqaligntwo}
C_1^{L} &=\frac{1}{2}g^2(g'^2y_LY_L\mp g^2\frac{1}{4}) &
C_0^{L} &=\frac{3}{16}g^4+g'^4y_L^2Y_L^2 \\
B_0^{R} &= g'^4y_R^2Y_L^2&
B_0^{L}&=g'^4y_L^2Y_R^2
       \end{eqaligntwo}
\begin{equation}
A_0^R=g'^4y_R^2Y_R^2
\end{equation}
\end{subequations}
This time, differently from Eqns. (\ref{55}-\ref{6.5}), 
the convention is that the upper index is for the initial lepton chirality.
So  $B_0^{R}$ indicates an initial state with 
a R lepton and a L quark, and so on. 
The upper (lower) sign refers to the $lq$ ($\bar{l}q$) case.

The energy and angular dependence of the resummed cross section can be found
as before:
\be\label{63}
\frac{d\sigma_{ij}}{d\cos\theta}=
\frac{N_c}{128\,\pi s}\frac{s^2}{t^2}
\left[
(A_0^R+C_0^L\pm C_1^L e^{-2\dl})(1+\cos\theta)^2
+4\,B_0^R+4\,B_0^L
\right]
\ee
We can neglect the terms $\propto g'^4$ in (\ref{62}), keeping the terms
$\propto g^2g'^2$ that are suppressed only by a factor
$\frac{g'^2}{g^2}=t_W^2$ and therefore not negligible in general, 
and thus obtaining:
\be\label{nuova}
\frac{d\sigma_{ij}}{d \cos \theta}=
 \alpha_W^2\,
\frac{N_c\,\pi}{2\, s}\frac{(1+\cos \theta)^2}
{(1-\cos \theta)^2}
\left(\frac{3}{16}+(-1)^{i+j}
(\mp\frac{1}{8} +\frac{1}{2} y_LY_Lt_W^2)
e^{-2\dl}\right)
\ee
where the  -
sign refers  to the $lq$ and the + sign to the $\bar{l}{q}$ case.

Although the BN violating corrections are regulated by the universal
Eqns. (\ref{risfin}), it should be clear from (\ref{risfin2}) that their
relative effect is dependent on the magnitude of
$\sigma_{12}^H$ and $\sigma_{11}^H$.
 In particular, in the case of left polarized beams one has:
\begin{subequations} \label{neutrino}
 \begin{eqaligntwo}\label{neutrino-a}
\frac{\Delta\sigma_{\bar{e}u}}{\sigma_{\bar{e}u}^H}\simeq 2.98\,\dl\quad & 
\frac{\Delta\sigma_{\bar{e}d}}{\sigma_{\bar{e}d}^H}\simeq
-0.75 \, \dl
&
\frac{\Delta\sigma_{eu}}{\sigma_{eu}^H}\simeq -0.84  \, \dl \quad &
\frac{\Delta\sigma_{ed}}{\sigma_{ed}^H}\simeq 5.4 \, \dl
\end{eqaligntwo}
\be\label{neutrino-b}
\frac{\Delta\sigma_{e \bar{\nu}_{\mu} }}{\sigma^H_{e \bar{\nu}_{\mu} }}
=
\frac{\Delta\sigma_{\nu_{\mu}\bar{e} }}{\sigma^H_{\nu_{\mu} \bar{e}}}
\simeq 10.6\;\dl
\ee
\end{subequations}

The magnitude of the effect in the electron-muon antineutrino scattering in
(\ref{neutrino-b})  is given by the $\propto g'^2 g^2$ terms in 
(\ref{nuova}). 
In fact, the cross sections for $\bar{e}u$ and $\bar{e}\nu_\mu$ are
equal in the $g'\to 0$ limit. However, the different hypercharges
for $\nu_\mu,u$ and the signs in formulas
(\ref{62},\ref{63}) conspire to produce a particularly small hard cross
section in the $\bar{e}\nu_\mu$ case.

Considering unpolarized beams, it is interesting to discuss the dependence
of the effect on the scattering angle. As we can see from
Fig. \ref{fig:tchannel}, there is no
effect for $\cos\theta=-1$, the reason being that the LL component does not
contribute to the cross section in this limit. The effect then increases
greatly with the angle. A cutoff on the maximum value of $\cos\theta$ is
necessary, since we always work in the limit $1-|\cos \theta| \gg M^2/s$.

The physical process proceeds via a proton; then, since
the effect has opposite signs for u and d quarks in (\ref{neutrino-b}), the
overall effect gets diminished.
For instance, taking the simplest picture of an hadron constituted only 
by  valence quarks, we obtain for polarized
left handed beams:

\be
\frac{\Delta\sigma_{ep}}{\sigma^H_{ep}}
\approx\frac{2\Delta\sigma_{eu}+\Delta\sigma_{ed}}
{2\sigma_{eu}+\sigma_{ed}}
\simeq -0.39\;  \dl\qquad\qquad
\frac{\Delta\sigma_{\bar{e}{p}}}{\sigma^H_{\bar{e} {p}}}
\approx 0.5\;\dl\qquad
\ee
This means that for hadron beams the relative effect is about one-half
that for lepton beams in Eq.(\ref{>?}).

\subsection{$q\bar{q}$ scattering}
In the $q\bar{q}$ case the overlap matrix
$O_H$ contains, besides s- channel and t - channel contributions, also
interference terms in the identical quarks case, and is characterized by the
additional presence of QCD contributions, reguated by the strong coupling
constant $\alpha_S(s)$, where $s\gg M^2$ denotes the hard process scale. 

In order to have a preliminary understanding, we neglect EW contributions to
$O_H$, and we note that EW BN violating corrections are absent of course in
the  gluon-gluon scattering cases, but also in the gluon - quark case
(see (\ref{44}) and discussion thereafter). In the example of 
pure $q_L\bar{q}_L$ s-channel annihilation on the other hand, 
we expect the possibility of having big relative effects at the parton
level, because of the hierarchy between $\sigma_{12}$ and $\sigma_{11}$. 
For instance, $\sigma_{u\bar{d}}^H$ vanishes in the limit considered here,
while $\sigma_{u\bar{u}}^H$ is of order $\alpha_S^2$, so that we obtain (in
the LL case):
\ba
\sigma_{u_L\bar{u}_L}&=&\sigma_{d_L\bar{d}_L}=
\frac{1}{2}\sigma^H_{u_L\bar{u}_L}\left(1+e^{-2\dl}\right)
\\
\sigma_{u_L\bar{d}_L}&=&\sigma_{d_L\bar{u}_L}=
\frac{1}{2}\sigma^H_{u_L\bar{u}_L}\left(1-e^{-2\dl}\right)
\ea
so that the corrections to, say, the cross section for  $u_L\bar{u}_L$ is of
the order of the cross section itself.
However, one should be cautious here because the observable effect with hadron
beams is not really large. Considering for instance the case
of valence quarks in $p\bar{p}$ collisions
we first take into
account that BN violating effects involve only left particles, so that the
unpolarized parton level cross sections are:
\begin{subequations} 
       \begin{eqalignno}
\sigma_{u\bar{u}} =\sigma_{d\bar{d}}
&=\sigma_{u_L\bar{u}_L}+\sigma_{u_R\bar{u}_R}
=\sigma_{u\bar{u}}^H\left(\frac{1+e^{-2\dl}}{4}+\frac{1}{2}\right)
\\
\sigma_{u\bar{d}}=\sigma_{d\bar{u}} 
&=\sigma_{u_L\bar{d}_L}
=\sigma_{u\bar{u}}^H\left(\frac{1-e^{-2\dl}}{4}\right)
       \end{eqalignno}
\end{subequations}
Next, a further suppression of the effect comes about because of a partial
cancellation between $u\bar{u}$ and  $u\bar{d}$ channels. 
In fact, since scattering occurs with probability
$\frac{5}{9}$ in a $u\bar{u}$ or  $d\bar{d}$ configuration and 
$\frac{4}{9}$ in a $u\bar{d}$ or $d\bar{u}$ configuration, we obtain:
\be
\sigma_{p\bar{p}}\approx\frac{5}{9}\sigma_{u\bar{u}}+
\frac{4}{9}\sigma_{u\bar{d}}=
\sigma_{p\bar{p}}^H\left(1+\frac{e^{-2\dl}-1}{20}\right)
\qquad ;\qquad
\frac{\Delta\sigma_{p\bar{p}}}{\sigma_{p\bar{p}}^H}\approx-\frac{1}{10}
\dl
\ee
and about twice as much in the pure LL channel. Therefore, the relative effect
is reduced to the {\it one percent} 
range at the TeV threshold.

\section{Outlook}
The outcome of this paper is that electroweak corrections become pretty large
at the TeV scale, even for inclusive processes, because of uncancelled 
double logarithmic enhancements involving the effective coupling $\dl$.
This leads to a sort of early unification within the Standard Model itself,
because strong and EW corrections are to be considered together much before
the respective running couplings become comparable.

\begin{figure}[htb]\setlength{\unitlength}{1cm}
\begin{picture}(12,8)
\put(3,1){\epsfig{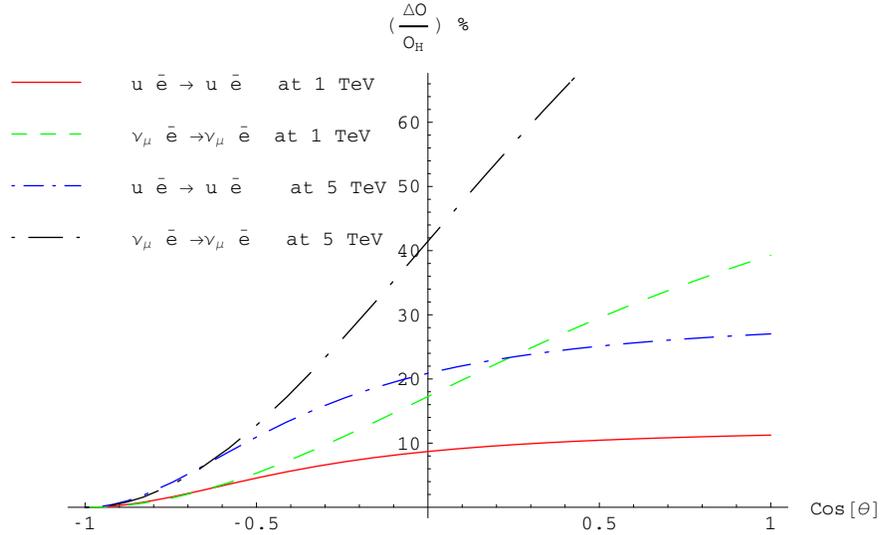}}
\end{picture}
\caption{\label{fig:tchannel}
Relative effects for $O=\frac{d\sigma}{d\cos\theta}
(\nu_\mu \bar{e})$ and $O=\frac{d\sigma}{d\cos\theta}
(u\bar{e})$
at $\sqrt{s}=$ 1 TeV 
and $\sqrt{s}=$ 5 TeV} 
\end{figure}

From a theoretical point of view, the effect above is due to both the 
nonabelian nature of the $SU(2)$ component of the standard model and to
symmetry breaking itself, which allows the initial states to be prepared
as abelian charges. It follows that all initial states carrying nontrivial
weak isospin are affected by the uncancelled double logs. For initial
fermions we find the following features:

(i)Only left-handed doublets are affected. This means that the BN violating
effect is strongly polarization dependent, and may lead to nontrivial angular
dependence in some cases (Fig. \ref{fig:tchannel}).

\begin{figure}[b]\setlength{\unitlength}{1cm}
\begin{picture}(12,8)
\put(3,0){\epsfig{file=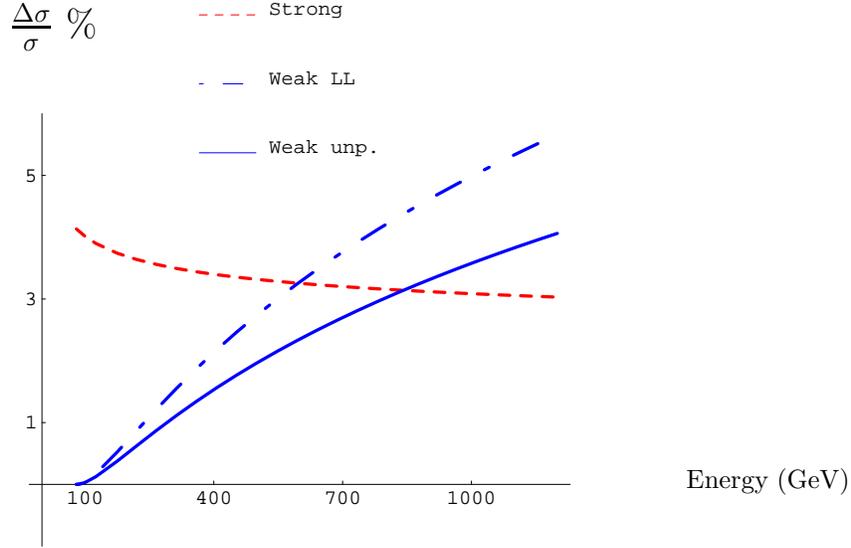,height=8cm}}
\put(3,7){\Large $\frac{\Delta\sigma}{\sigma}$ \%}
\put(12,1){Energy (GeV) }
\end{picture}
\caption{\label{had} Resummed double log EW corrections to $e^+e^-\to$ hadrons 
and strong corrections (dashed line) up to 3 loops. 
The dash-dotted line is
for a LL polarized beam, while the continuos line is for an 
unpolarized beam.}
\end{figure}

(ii)The effect is particularly important for purely leptonic beams, for which
its size is directly provided by $\dl$, which is about 7 \% at the TeV
threshold. For instance, in the case of $e_L^+e_L^-\to$ hadrons, despite
some reduction by exponentiation and running coupling effects, EW 
corrections are already 5.2 \%, compared to 3 \% strong corrections 
(Fig. \ref{had}).

(iii)There is a composite state reduction of the effect for hadron beams,
which act in the hard process as mixtures of partonic isospin states (recall
that the effect vanishes for a mixture with equal weights). The suppression
is of about 50 \% in the $lp,\bar{l}p$ scattering processes considered in 
Sec. 6.2. It is even stronger for hadron-hadron beams, due to both initial
isospin averaging and to QCD being flavor blind. Nevertheless, sizeable
effects in the {\it percent} range are still expected in the $p\bar{p}$
case (Sec. 6.3). The more relevant $pp$ case requires a detailed 
analysis, in which both structure functions and boson initiated processes
will presumably play a role.

Undoubtedly, quantitative estimates of the effects presented here are
needed for the planning of future accelerators. This implies not only
a more detailed analysis of Born cross sections, but also an extension
to subleading corrections which is far from being trivial \cite{cc2,fadin}.
What we learn here is that even at inclusive level we need to revise the
factorization theorems we are used to in QCD, in order to cope with fixed
initial flavor and to disentangle the enhanced SM corrections from new
physics.

\end{document}